\documentclass[11pt]{article}
\usepackage{latexsym,wrapfig,picinpar,picins}
\usepackage{amssymb}
\usepackage{graphicx}
\usepackage{mathrsfs}
\usepackage{epsfig,latexsym}
\usepackage{amsmath}
\usepackage{amssymb}
\usepackage{fancyref}
\usepackage{epsfig}
\usepackage{dcolumn}
\usepackage{CJK}
\usepackage{color}
\usepackage{picins}
\usepackage{pstricks}
\usepackage{epstopdf}
\usepackage{array}
\usepackage{hyperref}
\newcolumntype{C}{3cm}
\usepackage{subfigure}

\newtheorem{remark}{Remark}[section]

\newtheorem{theorem}{Theorem}[section]
\newtheorem{prop}{Proposition}[section]

\renewcommand{\theequation}{\arabic{section}.\arabic{equation}}

\newcommand{\F}{{\mathcal{F}}}

\newcommand{\D}{{\mathrm{d}}}

\renewcommand{\arraystretch}{.7}
\allowdisplaybreaks
\begin{document}
\begin{CJK}{GBK}{song}
%%%%%%%%%%%%%%%%%%%%%%%%%%%%%%%%%%%%%%%%%%%%%%%%%%%%%%%%%%%%%%%%%%%%%%%%%%
\title{\bf Pricing vulnerable options in a hybrid credit risk  model driven by Heston-Nandi GARCH processes\footnote{Gechun Liang
is at Department of Statistics, University of Warwick, Coventry CV4 7AL, UK.
Xingchun Wang is at the School of International Trade and Economics, University of International Business and Economics, Beijing 100029, China.
This study was supported by the National Natural Science Foundation of China (Nos. 11701084 and 11671084)  and Excellent Young Scholars Program in University of International Business and Economics (17YQ01). }
}

\author{GECHUN LIANG, XINGCHUN WANG\footnote{Correspondence address, Office 416, Qiuzhen Building, University of International Business and Economics, Beijing 100029, China.
Email: xchwangnk@aliyun.com; wangx@uibe.edu.cn}}
\date{}
\maketitle
\date{}
\maketitle

\begin{abstract}
This paper proposes a hybrid credit risk model, in closed form, to price vulnerable options with stochastic volatility. The distinctive features of the model are threefold. First, both the underlying and the option issuer's assets follow the Heston-Nandi GARCH model with their conditional variance being readily estimated and implemented solely on the basis of the observable prices in the market. Second, the model incorporates both idiosyncratic  and systematic risks into the asset dynamics of the underlying and the option issuer, as well as the intensity process. Finally, the explicit pricing formula of vulnerable options enables us to undertake the comparative statistics analysis. \\ \\
{\bf Keywords}: Vulnerable options; hybrid credit risk model; Heston-Nandi GARCH model; closed form formula.
\\ \\
{\bf JEL classification} : G13
 \end{abstract}

\section{Introduction}

 Vulnerable options refer to financial derivatives subject to  default risk of the option's issuers, and they are widely traded in over-the-counter (OTC) markets. {As of the first half of 2019, 3.9 trillion dollars (in terms of notional amounts) option contacts were traded in OTC markets\footnote{Resource: BIS, OTC derivatives statistics, \url{https://www.bis.org/statistics/derstats.htm}}.
 The bespoke nature and the flexibility in terms of product design have helped OTC markets to thrive. As opposed to exchange-traded derivatives for which products
are limited in tenor, size and strike ranges, OTC derivatives facilitate tailoring of transactions to meet specific end-users' needs.} In this paper, we study vulnerable options with stochastic volatility in a hybrid credit risk model driven by GARCH processes.

In order to study default risk of options, two types of models are widely used: \emph{structural models} and \emph{reduced-form models}.
Johnson and Stulz (1987) first investigate vulnerable options using the structural approach, where default happens when
the value of the option at maturity exceeds the value of the option issuer's assets, resulting in the failure of the option issuer to honor their obligation. This assumption is relaxed by
Klein (1996), where the option issuer could hold other liabilities having the same priority as the option. Vast majority of research focuses on the structural framework by taking into account of more factors such as stochastic interest rate, jump risk, stochastic volatility, stochastic default barriers, and multiple counterparties\footnote{A partial list of the studies on this topic includes Rich (1996), Klein and Inglis (1999), Klein and Inglis (2001), Cao and Wei (2001), Hui et al. (2003),  Liao and Huang (2005),  Kao (2006),  Liang and Ren (2007), Xu et al. (2012),  Tian et al. (2014),  Yang et al. (2014), Lee et al. (2016), Wang (2016), Wang et al. (2017), and Wang (2018).}. One attractive feature of the structural approach is its ability to explain default events via the structural variables such as asset dynamics.

As opposed to the structural approach, the reduced-form models are silent about why defaults happen and, instead, the dynamics of default are
exogenously given through a default rate, i.e. the default intensity. The latter approach is also called intensity approach.
In contrast to the reduced-form approach for bond pricing where the payoff is a fixed income, the payoff of vulnerable options is random, so it is more challenging in reduced-form models to obtain an explicit pricing formula of vulnerable options. There are relatively few results in this direction. To name a few, Hull and White (1995) impose an independence assumption to obtain a closed-form pricing formula of vulnerable options; Fard (2015) obtains a closed-form price for vulnerable options by assuming that the default intensity is captured by  a  mean-reverting Ornstein-Uhlenbeck process (so a negative intensity is allowed); Antonelli et al. (2020) employ a correlation expansion approach to
 provide an approximate evaluation of vulnerable option prices; and Wang (2017) obtains a closed-form solution for vulnerable options in a discrete-time GARCH framework.

In this paper, we consider vulnerable options in a hybrid credit risk model. The model will incorporate the attractive features of both
structural and reduced-form models. Hybrid credit risk models were initiated by  Madan and Unal (2000), who investigate the pricing issue of risky debt in a hazard rate model with two factors being the values of the firm's assets and the interest rate.
Bakshi et al. (2006) further work under a
reduced-form model based on Vasicek-type state variables, such as
leverage, book-to-market and equity-volatility.
Gu et al. (2014)  consider a new type of reduced-form
model that incorporates the impacts of  observable trigger events
 as well as economic environment on corporate defaults.
Boudreault et al. (2014) measure how contagion affects default time and recovery rates
in a hybrid model, where both the default probability and the recovery rate are functions
of the firm's leverage ratio. However, the above mentioned studies mainly focus on risky debt or credit derivatives. The aim of the current paper is to propose a hybrid credit risk model for vulnerable options with the aforementioned reduced-form model as a special case of the proposed hybrid credit risk model.

In our model, the dynamics of the underlying and the option issuer's assets follow the Heston-Nandi GARCH processes to incorporate stochastic volatility. As pointed out in Heston and Nandi (2000) and Hsieh and Ritchken (2005), the continuous time stochastic volatility models are difficult to implement and test, while GARCH models have inherent advantages that the volatility is readily observable from the history of asset prices. We assume that the asset values of both underlying and option issuer
are exposed to idiosyncratic and systematic risks.
Furthermore, we also allow the intensity process to be driven by idiosyncratic shocks of the issuer and systematic shocks of the market. Thus, the systematic risk factor correlates all the underlying processes in the proposed hybrid model.

Under this framework, we obtain an explicit pricing formula of vulnerable options based on the explicit expression of the joint  generating function and the change of measure technique. The joint generating function (see Proposition 2.1) generalizes the generating function for a single stock case in Heston and Nandi (2000) to a multidimensional case including the underlying stock, the issuer's assets and the intensity process.
Finally, we undertake comparative statistics analysis to investigate the effects of default risk on the option prices, and compare them with the default-free option prices and the ones obtained in the reduced-form model. One of the striking features is that the option prices increase with
the sensitivity of the issuer's assets to systematic risk, albeit a higher value of the sensitivity means that the issuer's assets are more risky, resulting in a higher possibility of default. This is because a larger value of sensitivity also means the underlying
asset and the issuer's assets are more likely to be correlated, which in turn makes option issuers less likely to default when  call options end in the money, yielding a higher option price consequently.

The remainder of this paper is organized as follows. In the coming section, we focus on the hybrid credit risk model and the derivation of the explicit pricing formulae. Section 3 is devoted to numerical results.
Finally, Section 4 summarizes and concludes the paper.  The detailed proofs are shown in the appendix.

\section{The hybrid credit risk model}
 \setcounter{equation}{0}

In this section, we propose a hybrid credit risk model to price vulnerable options. An explicit pricing formula of vulnerable options is derived based on the change of measure technique and the explicit expression of the joint characteristic function of underlying processes.

\subsection{The market}

Let $Q$ be a risk neutral probability measure on a filtered probability space $(\Omega,\F,(\F_t)_{t\geq0},Q)$. Consider a market with the systematic risk factor modelled by the market index $M(t)$, whose dynamics follow the Heston-Nandi GARCH process,
that is,
\begin{align}\label{dynamics of M under Q}
      \left\{\begin{array}{ll}
            \ln M(t) =\ln M(t-1)+r-\frac{1}{2}h_m(t)+\sqrt{h_m(t)}Z_m(t),\\
           h_m(t)= w_m+b_m h_m(t-1)+a_m\Big(Z_m(t-1)-c_m\sqrt{h_m(t-1)}\Big)^2,
     \end{array}\right.
\end{align}
where $r$ is the continuously compounded interest rate for the time interval $[t-1,t]$, and  $Z_m(t)$ is a standard normal random variable. The conditional variance $h_m(t)$ of the log return between $t-1$ and $t$ is known from the information set at time $t-1$, so it can be readily estimated and implemented solely on the basis of the observables. In the driving noise term of $h_m(t)$, the constant $a_m$ determines the kurtosis of the noise, and the constant $c_m$ results in asymmetric influence of the noise $Z_m(t-1)$.

It has been shown in Heston and Nandi (2000) that the continuous time limit of the conditional variance $h_m(t)$ is a square-root diffusion process corresponding to the continuous time Heston    stochastic volatility model. On the other hand, it is clear that the discounted price of the market index is a martingale under $Q$. Indeed, we have
\begin{eqnarray*}
E_{t-1}\Big[M(t)\Big]&=&E_{t-1}\Big[M(t-1)e^{r-\frac{1}{2}h_m(t)+\sqrt{h_m(t)}Z_m(t)}\Big]\nonumber\\
&=&M(t-1)e^rE_{t-1}\Big[e^{-\frac{1}{2}h_m(t)+\sqrt{h_m(t)}Z_m(t)}\Big]\nonumber\\
&=&M(t-1)e^r,
\end{eqnarray*}
where in the last equality we have used the fact that $h_m(t)$ is known given the information at time $t-1$ and $Z_m(t)$ is a standard normal variable.

Consider a stock in this market. Its return is affected by not only the systematic risk factor via $Z_m(t)$ but also the idiosyncratic risk factor via an independent normal random variable $Z_s(t)$. Hence, the stock price $S(t)$ under $Q$ is  driven by the process
\begin{align}\label{dynamics of S under Q}
     \left\{\begin{array}{ll}
          \ln S(t)  = \ln S(t-1)+r-\frac{1}{2}h_s(t)+\sqrt{h_s(t)}Z_s(t)
              -\frac{1}{2}\beta_s^2h_m(t)+\beta_s\sqrt{h_m(t)}Z_m(t),\\
          h_s(t)=w_s+b_s h_s(t-1)+a_s\Big(Z_s(t-1)-c_s\sqrt{h_s(t-1)}\Big)^2,
      \end{array}\right.
\end{align}
where the constant $\beta_s$
captures the sensitivity of the stock price to systematic risk. Since $h_m(t)$ and $h_s(t)$ are known given the information at time $t-1$, the independence assumption between $Z_m(t)$ and $Z_s(t)$ implies that the discounted value of $S(t)$ is also a martingale under $Q$,
\begin{eqnarray*}
E_{t-1}\Big[\frac{S(t)}{S(t-1)}\Big]&=&E_{t-1}\Big[e^{r-\frac{1}{2}h_s(t)+\sqrt{h_s(t)}Z_s(t)-\frac{1}{2}\beta_s^2h_m(t)+\beta_s\sqrt{h_m(t)}Z_m(t) }\Big]\nonumber\\
&=&e^rE_{t-1}\Big[e^{-\frac{1}{2}h_s(t)+\sqrt{h_s(t)}Z_s(t)}\Big] E_{t-1}\Big[e^{-\frac{1}{2}\beta_s^2h_m(t)+\beta_s\sqrt{h_m(t)}Z_m(t)}\Big]         \nonumber\\
&=&e^r.
\end{eqnarray*}

We consider a European call option written on the stock with strike price $K$ and maturity $T$, so its risk neutral price is given by $E[e^{-r T}(S(T)-K)^+]$
\emph{if the option issuer does not default during the contract period and is able to honor their obligation}. Note that under the above GARCH framework, Heston and Nandi (2000) derived an explicit pricing formula for the European call option using the characteristic function of $S(t)$ (see section 2 therein).

\subsection{The vulnerable option with credit value adjustment}

When the options are traded in OTC markets, the holders may face the potential default risk that the issuers are not able to deliver the promised payoff. We model the default risk in a hybrid model. To this end, let $N(t)$ be a doubly stochastic Poisson process (Cox process) with intensity $\Lambda(t)$, and
$\tau$ be its first jump time which can be regarded as the arrival time of the default trigger event as in Gu et al. (2014). A loss given default (LGD) will occur when the trigger event arrives, and it is given by a constant $L$.
Furthermore, assume that the option issuer would recover from the trigger event if the value of the issuer's assets is larger than the LGD $L$. Hence, default occurs only when the trigger event occurs and the value of the issuer's assets at the arrival time of the trigger event falls below the LGD $L$.

Next, we model the option issuer's assets $V(t)$ and the Cox process' intensity $\Lambda(t)$.
Assume that the return of the issuer's assets
is also affected by both the systematic and idiosyncratic risks and its dynamics follow
\begin{eqnarray}\label{dynamics of V}
 \left\{\begin{array}{ll}
  \ln V(t)  = \ln V(t-1)+r-\frac{1}{2}h_v(t)+\sqrt{h_v(t)}Z_v(t)-\frac{1}{2}\beta_v^2h_m(t)+\beta_v\sqrt{h_m(t)}Z_m(t),\\
  h_v(t)= w_v+b_v h_v(t-1)+a_v\Big(Z_v(t-1)-c_v\sqrt{h_v(t-1)}\Big)^2,
  \end{array}\right.
\end{eqnarray}
where $Z_v(t)$ is a standard normal variable independent of $Z_s(t)$ and $Z_m(t)$.
Note that $Z_m(t)$ captures the systematic risk, and $Z_s(t)$ and $Z_v(t)$ represent
the idiosyncratic risks of the underlying asset and the issuer's assets, respectively.
Similarly to $\beta_s$ in (\ref{dynamics of S under Q}), $\beta_v$
captures the sensitivity of the issuer's assets to the systematic risk.
As for the intensity process $\Lambda(t)$, we assume that it is driven by $Z_v(t)$ and $Z_m(t)$, the driving noise faced by the issuer. Specifically, the dynamics of $\Lambda(t)$ are given by
\begin{align}\label{intensity}
    \Lambda(t+1)= w_{\lambda}+b_{\lambda} \Lambda(t)+a_\lambda(Z_{m}(t))^2+    c_\lambda(Z_{v}(t))^2.
\end{align}
%where $Z_{\lambda}(t)$ is a standard normal variable under $Q$, independent of $Z_m(t)$, $Z_s(t)$ and $Z_v(t)$,
All the parameters are  non-negative to
 ensure that the intensity is non-negative.

We are now in a position to present the hybrid credit risk model for the valuation of vulnerable options. To take account of the issuer's default risk, we model the difference between the default-free value and the true value of the European option as follows:
When $j-1<\tau\leq j,\ V(j)<L$, i.e. the trigger event occurs between $(j-1,j]$ and the issuer's asset value falls below the LGD, suppose the option holder will then only receive $\alpha V(j)/L$ proportion of the nominal payoff $(S(T)-K)^+$ at the maturity $T$, where the constant $\alpha\in[0,1]$ represents the recovery rate and $(1-\alpha)V(j)$ represents the deadweight costs associated with the bankruptcy. Hence, the expected value of the credit value adjustment (i.e. the difference between the default-free value and the true value) is
$$E\left[e^{-rT}(1-\frac{\alpha V(j)}{L})(S(T)-K)^+\right].$$
conditional on the event $\{j-1<\tau\leq j, V(i)<L\}$.

The price of the vulnerable option at time $0$ is therefore given by
\begin{eqnarray}\label{C}
C_{0}&=&E\Big[e^{-r T}(S(T)-K)^+\Big]-\sum_{j=1}^{T}E\Big[e^{-r T}I(j-1<\tau\leq j,\ V(j)<L)  (1-  \frac{\alpha  V(j)}{L}) (S(T)-K)^+ \Big]\nonumber\\
&=&E\Big[e^{-r T}(S(T)-K)^+\Big]-\sum_{j=1}^{T}E\Big[e^{-r T}I(j-1<\tau\leq j,\ V(j)<L)  (S(T)-K)^+ \Big]\nonumber\\
&&+\ \  \sum_{j=1}^{T}E\Big[e^{-r T}I(j-1<\tau\leq j,\ V(j)<L)   \frac{\alpha V(j)}{L} (S(T)-K)^+ \Big],
%&:=& A_1+A_2+A_3,
\end{eqnarray}
where $I(\cdot)$ is the indicator function. The first term in (\ref{C}) is the  default-free value, the second term represents the costs when default occurs, and the last term is the recovery value from the default. Note that $I(j-1<\tau\leq j)=I(j-1<\tau)-I(j<\tau)$, so the last two terms in (\ref{C}) simplify to
\begin{eqnarray*}
&&E\Big[I(j-1<\tau\leq j,\ V(j)<L)  (S(T)-K)^+ \Big]\nonumber\\
&=& E\Big[I(j-1<\tau\leq j)I( V(j)<L)  (S(T)-K)^+ \Big]\nonumber\\
&=&- E\Big[I(j<\tau)I( V(j)<L)  (S(T)-K)^+ \Big] +E\Big[I(j-1<\tau)I( V(j)<L)  (S(T)-K)^+ \Big]\nonumber\\
&=&- E\Big[e^{-\sum_{k=1}^{j} \Lambda(k)}I( V(j)<L)  (S(T)-K)^+ \Big] + E\Big[e^{-\sum_{k=1}^{j-1} \Lambda(k)}I( V(j)<L)  (S(T)-K)^+ \Big],
\end{eqnarray*}
and
\begin{eqnarray*}
&&E\Big[I(j-1<\tau\leq j,\ V(j)<L)  \frac{\alpha V(j)}{L}  (S(T)-K)^+ \Big]\nonumber\\
&=& E\Big[I(j-1<\tau\leq j)I( V(j)<L)  \frac{\alpha V(j)}{L}  (S(T)-K)^+ \Big]\nonumber\\
&=& -E\Big[I(j<\tau)I( V(j)<L)  \frac{\alpha V(j)}{L}  (S(T)-K)^+ \Big] \\
&&+ E\Big[I(j-1<\tau)I( V(j)<L)  \frac{\alpha V(j)}{L}  (S(T)-K)^+ \Big]\nonumber\\
&=& -E\Big[e^{-\sum_{k=1}^{j} \Lambda(k)}I( V(j)<L) \frac{\alpha V(j)}{L}   (S(T)-K)^+ \Big]\\
&&+ E\Big[e^{-\sum_{k=1}^{j-1} \Lambda(k)}I( V(j)<L)  \frac{\alpha V(j)}{L}  (S(T)-K)^+ \Big].
\end{eqnarray*}
In turn, we have
\begin{eqnarray}\label{CC}
C_{0}
&=&e^{-r T}\Big(E\Big[(S(T)-K)^+\Big]\nonumber\\
&& +\sum_{j=1}^{T}  E\Big[e^{-\sum_{k=1}^{j} \Lambda(k)}I( V(j)<L)  (S(T)-K)^+ \Big]\nonumber\\
&& -\sum_{j=1}^{T}  E\Big[e^{-\sum_{k=1}^{j-1} \Lambda(k)}I( V(j)<L)  (S(T)-K)^+ \Big]\nonumber\\
&& -\sum_{j=1}^{T} E\Big[e^{-\sum_{k=1}^{j} \Lambda(k)}I( V(j)<L) \frac{\alpha V(j)}{L}   (S(T)-K)^+ \Big]\nonumber\\
&&+\sum_{j=1}^{T} E\Big[e^{-\sum_{k=1}^{j-1} \Lambda(k)}I( V(j)<L)  \frac{\alpha V(j)}{L}  (S(T)-K)^+ \Big]\Big).
\end{eqnarray}
{
\begin{remark}
In the proposed framework, the correlation coefficient between the underlying asset and the issuer's assets is given by
\begin{eqnarray*}
\frac{{\rm Cov}_t(\ln\frac{ S(t+1)}{S(t)}, \ln\frac{ V(t+1)}{V(t)})}{\sqrt{{\rm Var}_t(\ln\frac{ S(t+1)}{S(t)})}\sqrt{{\rm Var}_t(\ln\frac{ V(t+1)}{V(t)}})}
&=&\frac{{\rm Cov}_t(\beta_s\sqrt{h_m(t+1)}Z_m(t+1), \beta_v\sqrt{h_m(t+1)}Z_m(t+1))}{\sqrt{h_s(t+1)+\beta_s^2h_m(t+1)}\sqrt{h_v(t+1)+\beta_v^2h_m(t+1)}}\nonumber\\
&=&\frac{\beta_s\beta_vh_m(t+1)}{\sqrt{h_s(t+1)+\beta_s^2h_m(t+1)}\sqrt{h_v(t+1)+\beta_v^2h_m(t+1)}}.
\end{eqnarray*}
When $\beta_s=0$ or $\beta_v=0$,  the underlying asset and the issuer's assets are not correlated with each other.
On the other hand, when $h_s(t+1)\equiv 0$ (i.e. $w_s=b_s=a_s=0$) and $h_v(t+1)\equiv 0$ (i.e. $w_v=b_v=a_v=0$), both the underlying asset and the issuer's assets are only driven by $Z_m(t)$, and the correlation coefficient becomes to be $\pm 1$. In this sense, we can view $Z_m(t)$ as a common risk factor in the returns on the underlying asset and the issuer's assets, and the issuer could hedge the option position by directly trading the underlying asset. Thus, $Z_m(t)$ could represent not only the systematic risk factor (though such an interpretation is the most typical example).
\end{remark}
}

\subsection{The explicit pricing formula}
In order to obtain an explicit pricing formula for vulnerable options in the proposed framework, we first derive
the joint conditional generating function of the underlying processes. To this end, let $f(t;\phi_1,\phi_2,\phi_3,\phi_4)$ denote the conditional generating function given below,
$$f(t;\phi_1,\phi_2,\phi_3,\phi_4)=E_{t}\Big[\exp\Big\{\phi_1 \ln S(T) +\phi_2\ln V(j) +\phi_3 \Lambda(j) +\phi_4\sum_{k=1}^{j-1} \Lambda(k)  \Big\}\Big],$$
where $j\leq T$ and $0\leq t\leq T$. Specially, $f(t;\phi_1,0,0,0)$ is the conditional generating function of the underlying asset and can be used to derive
the default-free value of the European option as in Heston and Nandi (2000). In addition, $f(t;\phi_1,0,\phi_3,\phi_4)$
can be employed to obtain the closed-form pricing formula of vulnerable options in the reduced-form models, which is a special case of the proposed hybrid credit risk model (see section 2.4).

In the proposed framework, the explicit expression of $f(t;\phi_1,\phi_2,\phi_3,\phi_4)$ is available and given in the following proposition.
\begin{prop}\label{generating function}
The  conditional generating function  has the following form\footnote{For convenience, we use the more parsimonious notation $f(t)$ to indicate $  f(t;\phi_1,\phi_2,\phi_3,\phi_4)$, and similarly for $A_i(t)$ and $B_i(t)$.}
\begin{eqnarray}
f(t)
&=&\exp\Big\{\phi_2\ln V(j) +\phi_3 \Lambda(j) +\phi_4\sum_{k=1}^{j-1} \Lambda(k)  +\phi_1\ln S(t)+A_0(t)\nonumber\\
 && \ \ \ \ \ \ \ \ \ \ \ +A_1(t)h_m(t+1)+A_2(t)h_s(t+1)\Big\},
\end{eqnarray}
for $j\leq t\leq T$, where $A_0(t)$, $A_1(t)$ and $A_2(t)$ ($j\leq t\leq T$) are defined recursively with terminal conditions $A_0(T)=A_1(T)=A_2(T)=0$ by the following expressions
\begin{align*}
A_0(t)=&  \phi_1 r+A_0(t+1)+ w_m A_1(t+1)+w_sA_2(t+1)-\frac{1}{2}\ln(1-2a_mA_1(t+1))\\
&-\frac{1}{2}\ln(1-2a_sA_2(t+1)),\\
A_1(t)=& b_mA_1(t+1)-\frac{1}{2}\phi_1\beta_s^2
+\phi_1\beta_sc_m-\frac{1}{2}c_m^2+\frac{\frac{1}{2}(\phi_1\beta_s-c_m)^2}{1-2a_mA_1(t+1)},\\
A_2(t)=& b_sA_2(t+1)-\frac{1}{2}\phi_1+\phi_1c_s-\frac{1}{2}c_s^2+\frac{\frac{1}{2}(\phi_1-c_s)^2}{1-2a_sA_2(t+1)}.
\end{align*}
For $t< j$,
\begin{eqnarray}
f(t)
&=&\exp\Big\{\phi_2\ln V(t)+\phi_4\sum_{k=1}^{t} \Lambda(k)  +\phi_1\ln S(t)+B_0(t)\nonumber\\
 && \ \ \ \ \ \ +B_1(t)h_m(t+1)+B_2(t)h_s(t+1)+B_3(t)h_v(t+1)+B_4(t+1)  \Lambda(t+1) \Big\},
\end{eqnarray}
where $B_{k}(t),\ k=0,1,2,3,4$ ($t< j$) can be obtained recursively by the following expressions
\begin{eqnarray*}
B_0(t)&=&B_0(t+1)+   (\phi_2+\phi_1) r +w_m B_1(t+1)+w_sB_2(t+1)+w_vB_3(t+1)+w_{\lambda}B_4(t+1)\\
&&\ \ \ -\frac{1}{2}\ln (1-2 (a_mB_1(t+1)+a_\lambda B_4(t+1) ))-\frac{1}{2}\ln (1-2 a_sB_2(t+1)))\\
&&\ \ \  -\frac{1}{2}\ln (1-2 (a_vB_3(t+1)+c_\lambda B_4(t+1) )),\\
B_1(t)&=&b_mB_1(t+1)-\frac{1}{2}\phi_2\beta_v^2-\frac{1}{2}\phi_1\beta_s^2+ a_mc_m^2B_1(t+1)\\
&&\ \ \ +\frac{2(a_mc_mB_1(t+1)-(\phi_2\beta_v+ \phi_1\beta_s )/2)^2 }{1-2(a_mB_1(t+1)+a_\lambda B_4(t+1) )},\\
B_2(t)&=&b_sB_2(t+1)-\frac{1}{2}\phi_1 + a_sc_s^2B_2(t+1)+\frac{2(a_sc_sB_2(t+1)- \phi_1/2)^2 }{1-2a_sB_2(t+1)},\\
B_3(t)&=&b_vB_3(t+1)-\frac{1}{2}\phi_2+ a_vc_v^2B_3(t+1)+\frac{2(a_vc_vB_3(t+1)- \phi_2/2)^2 }{1-2(a_vB_3(t+1)+c_\lambda B_4(t+1) )},\\
B_4(t)&=& b_{\lambda}B_4(t+1)+\phi_4.
\end{eqnarray*}
Moreover, terminal conditions  $B_{k}(j-1),\ k=0,1,2,3,4$  are determined by $A_0(j)$, $A_1(j)$ and $A_2(j)$ as follows:
\begin{eqnarray*}
B_0(j-1)&=&A_0(j)+   (\phi_2+\phi_1) r +w_m A_1(j)+w_sA_2(j)-\frac{1}{2}\ln (1-2 a_mA_1(j))-\frac{1}{2}\ln (1-2 a_sA_2(j)),\\
B_1(j-1)&=&b_mA_1(j)-\frac{1}{2}\phi_2\beta_v^2-\frac{1}{2}\phi_1\beta_s^2+ a_mc_m^2A_1(j) +\frac{2(a_mc_mA_1(j)-(\phi_2\beta_v+ \phi_1\beta_s )/2)^2 }{1-2a_mA_1(j)},\\
B_2(j-1)&=&b_sA_2(j)-\frac{1}{2}\phi_1 + a_sc_s^2A_2(j)+\frac{2(a_sc_sA_2(j)- \phi_1/2)^2 }{1-2a_sA_2(j)},\\
B_3(j-1)&=&-\frac{1}{2}\phi_2+\frac{1}{2}\phi_2^2,\\
B_4(j-1)&=& \phi_3.
\end{eqnarray*}
\end{prop}
Proof. See the appendix.\\

We are ready to obtain the closed form pricing formula of the vulnerable option price in (\ref{C}).
\begin{theorem}\label{option prices}
The price of the vulnerable European call option with strike price $K$ and maturity $T$ is given by
\begin{eqnarray*}\label{prices}
C_{0}&=& e^{-r T}\Big(  \frac{1}{2}f(0;1,0,0,0)+\frac{1}{\pi}\int_{0}^\infty \textrm{Re}\Big[\frac{e^{-i \phi_1 \ln K}f(0;1+i\phi_1,0,0,0)}{ i\phi_1 }\Big]\D \phi_1\nonumber\\
&&\ \ \ \ \ \ \ \ \ \ \ \ \     - \frac{K}{2}-\frac{K}{\pi}\int_{0}^\infty \textrm{Re}\Big[\frac{e^{-i \phi_1 \ln K}f(0;i\phi_1,0,0,0)}{ i\phi_1 }\Big]\D \phi_1  \nonumber\\
&&\ \ \ \ \ \ \ \ \ \ \ \ \  +\sum_{j=1}^{T} \Big( \Pi_{1,j} -  \Pi_{3,j}-\frac{\alpha}{L} \Pi_{5,j}+\frac{\alpha}{L} \Pi_{7,j}  -K(\Pi_{2,j} -\Pi_{4,j} - \frac{\alpha}{L} \Pi_{6,j}+\frac{\alpha}{L} \Pi_{8,j})\Big)\Big),
\end{eqnarray*}
where $\Pi_{j,1}$-$\Pi_{j,8}$ are given in (\ref{pi1})-(\ref{pi8}).
\end{theorem}
Proof. See the appendix.

\subsection{Comparison with reduced-form models}

Reduced-form models can be seen as a special case of the proposed hybrid credit risk model. % To connect with the structural model, we discard the trigger event $\tau$ and instead
%compare the LGD $L$ with the asset value $V(j)$ at each integer time $j$. Thus, default happens at the first integer time $j$ such that $V(j)$ falls below $L$. This corresponds to the first-passage-time structural model, and the price is given by
%\begin{eqnarray}\label{CS}
%C_{0}^S
%&=&E\Big[e^{-r T}(S(T)-K)^+\Big]-\sum_{j=1}^{T}E\Big[e^{-r T}\prod _{i=1}^{j-1}I(V(i)\geq L)I(V(j)<L)  (S(T)-K)^+ \Big]\nonumber\\
%&&+\ \  \sum_{j=1}^{T}E\Big[e^{-r T}\prod _{i=1}^{j-1}I(V(i)\geq L)I(V(j)<L)\frac{\alpha V(j)}{L} (S(T)-K)^+ \Big],
%%&:=& A_1+A_2+A_3,
%\end{eqnarray}
%with the convention $\prod_{i=1}^0\equiv 1$.
%
%On the other hand,
To connect with the reduced-form model, we discard the LGD $L$ and only check the default trigger event $\tau$. Hence, the price in the reduced-form model is given by
\begin{eqnarray}\label{CR}
{C}_{0}^R
&=&E\Big[e^{-r T}(S(T)-K)^+\Big]-(1- \alpha)  \sum_{j=1}^{T}E\Big[e^{-r T}I(j-1<\tau\leq j)   (S(T)-K)^+ \Big].
\end{eqnarray}

The vulnerable option price in (\ref{CR}) is given in the following theorem.
\begin{theorem}\label{option prices}
 In the reduced-form model, the price of the vulnerable European call option with strike price $K$ and maturity $T$ is given by
\begin{eqnarray*}
C^R_{0}&=& e^{-r T}\Big(  \frac{1}{2}f(0;1,0,0,0)+\frac{1}{\pi}\int_{0}^\infty \textrm{Re}\Big[\frac{e^{-i \phi_1 \ln K}f(0;1+i\phi_1,0,0,0)}{ i\phi_1 }\Big]\D \phi_1\nonumber\\
&&\ \ \ \ \ \ \ \ \ \ \ \ \     - \frac{K}{2}-\frac{K}{\pi}\int_{0}^\infty \textrm{Re}\Big[\frac{e^{-i \phi_1 \ln K}f(0;i\phi_1,0,0,0)}{ i\phi_1 }\Big]\D \phi_1  -  (1-\alpha)\sum_{j=1}^{T} (\bar{\Pi}_{j,1}-\bar{\Pi}_{j,2})\Big),
\end{eqnarray*}
where
\begin{eqnarray*}
\bar{\Pi}_{j,1}&=&\frac{1}{2}f(0;1,0,0,-1)+\frac{1}{\pi}\int_{0}^\infty \textrm{Re}\Big[\frac{e^{-i \phi_1 \ln K}f(0;1+i\phi_1,0,0,-1)}{ i\phi_1 }\Big]\D \phi_1\nonumber\\
&&\ \ \ \ \ \ \ \ \ \ \ \ \     - \frac{K}{2}f(0;0,0,0,-1)-\frac{K}{\pi}\int_{0}^\infty \textrm{Re}\Big[\frac{e^{-i \phi_1 \ln K}f(0;i\phi_1,0,0,-1)}{ i\phi_1 }\Big]\D \phi_1,\nonumber\\
\bar{\Pi}_{j,2}&=&\frac{1}{2}f(0;1,0,-1,-1)+\frac{1}{\pi}\int_{0}^\infty \textrm{Re}\Big[\frac{e^{-i \phi_1 \ln K}f(0;1+i\phi_1,0,-1,-1)}{ i\phi_1 }\Big]\D \phi_1\nonumber\\
&&\ \ \ \ \ \ \ \ \ \ \ \ \     - \frac{K}{2}f(0;0,0,-1,-1)-\frac{K}{\pi}\int_{0}^\infty \textrm{Re}\Big[\frac{e^{-i \phi_1 \ln K}f(0;i\phi_1,0,-1,-1)}{ i\phi_1 }\Big]\D \phi_1.
\end{eqnarray*}
\end{theorem}

The main difference between the  proposed hybrid model and the above reduced-form model is that we discard the LGD $L$ (so we are silent about why the issuer defaults).
In section 3, we will compare the proposed hybrid model with the above reduced-form model numerically.

\section{Numerical Results}

In this section, we undertake comparative statistics analysis for the vulnerable option prices in the proposed hybrid credit risk  model. For comparison purpose, we also report the values of the corresponding European options without default risk and vulnerable options in the reduced-form model in section 2.4.
{In particular, the default premiums, i.e. the price differences between
the vanilla European options and
the above two vulnerable option prices, are illustrated.
}

%In order to show the impacts of default risk, we illustrate the price difference

In order to calculate the prices, we use the values of the parameters listed in Table \ref{T:para1}.
These parameter values in the dynamics of the market index and the underlying asset are also used in
Su and Wang (2019), and they are estimated  based on  the daily closing values of the S\&P 500 index and its five largest stocks
for the period from January $3$, $2000$ to  May $31$, $2018$. In addition, the initial variance values are set to be squared stationary volatilities.
The parameter values in the intensity process can produce average cumulative default rates for corporate bonds with a credit rating of B, i.e.,
 $5.33\%$, $16.19\%$, $25.89\%$ and $34.47\%$ for
 $1.0$, $3.0$, $5.0$ and $7.0$ years, respectively (see, e.g., Table 22.1 in Hull (2012)).
For simplicity, the parameter values in the issuer's asset dynamics are set to be the same as those in the underlying asset.

\begin{table}
\centering
\renewcommand{\arraystretch}{1.2}
\makeatletter
\def\@captype{table}
\setlength{\abovecaptionskip}{0pt}
\setlength{\belowcaptionskip}{10pt}
\makeatother \caption{\small Parameter Values\label{T:para1}}
\begin{tabular}{p{8.0cm} c c c c c c}

  \hline
Parameters in the market index dynamics   &   &\\
Initial price&   $M(0)=$1 &\\
 Initial variance &  $h_m(0)=$3.27E-02&\\
Parameters governing variance processes&   $w_m=$ 7.10E-13       &   $b_m=$7.67E-01   \\
                                    &    $a_m=$2.99E-06      &      $c_m=$2.65E+02         \\
\hline
 Parameters in the underlying asset dynamics   &   &\\
Initial price&   $S(0)=$1 &\\
 Initial variance &  $h_s(0)+\beta_s^2h_m(0)=$1.22E-01&  $\beta_s=$1.15\\
Parameters governing variance processes&   $w_s=$9.79E-07        &   $b_s=$9.55E-01   \\
                                    &    $a_s=$3.71E-06        &      $c_s=$9.01E+01         \\
\hline
 Parameters in the default intensity &   & \\
Initial intensity & $\lambda(0)=$1.275E-06&\\
Parameters governing default intensities& $w_{\lambda}= $8.637E-07 &$a_{\lambda}=$1.372E-10\\
&  $b_{\lambda}=$9.949E-01&$c_{\lambda}=$1.372E-10\\
 \hline
  Parameters in  the value of the issuer's assets    &   &\\
Initial price&   $V(0)=$1 &\\
 Initial variance &  $h_v(0)+\beta_v^2h_m(0)=$1.22E-01&  $\beta_v=$1.15\\
Parameters governing variance processes&   $w_v=$9.79E-07      &   $b_v=$9.55E-01  \\
                                    &    $a_v=$3.71E-06       &      $c_v=$9.01E+01         \\
\hline
Other parameters    &  & \\
Interest rate & $r=0.05$&\\
  Strike price & $K=1$&\\
Maturity &$T=2.0$&\\
Recovery rate&$\alpha=0.50$&\\
Caused loss & $L=90$\\
  \hline
  \end{tabular}
\end{table}

\begin{figure}
\begin{center}
\makeatletter
\def\@captype{figure}
\makeatother
     \includegraphics[width=4.0in]{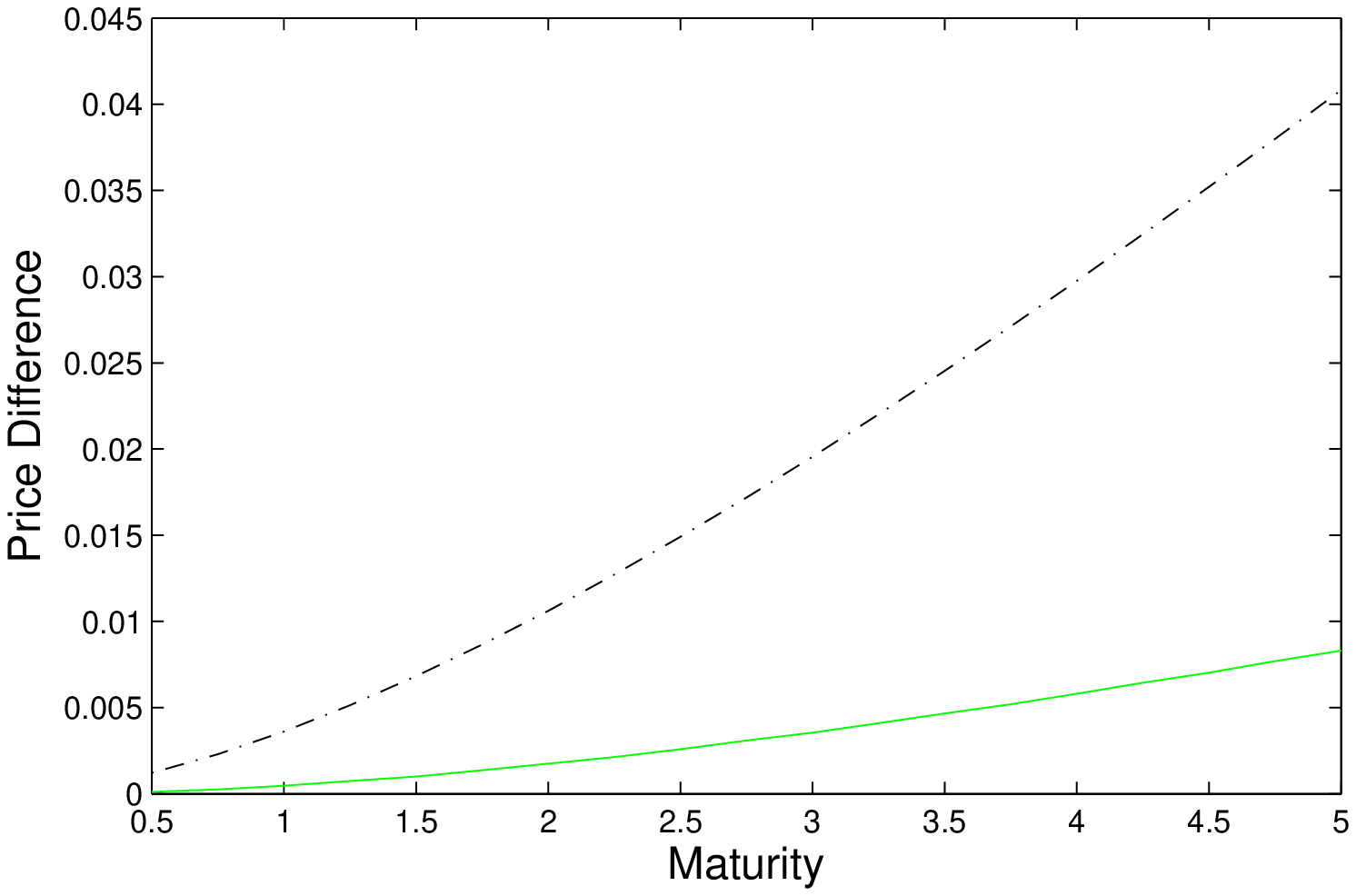}
     \caption{  Option price differences against maturities.  The solid line corresponds to
     the price difference between the default-free model and  the proposed hybrid model, and the dot-dashed line
     corresponds to the price difference between the default-free model and  the reduced-form model.    \label{T}}
\end{center}
\end{figure}

\begin{figure}
\begin{center}
\makeatletter
\def\@captype{figure}
\makeatother
     \includegraphics[width=4.0in]{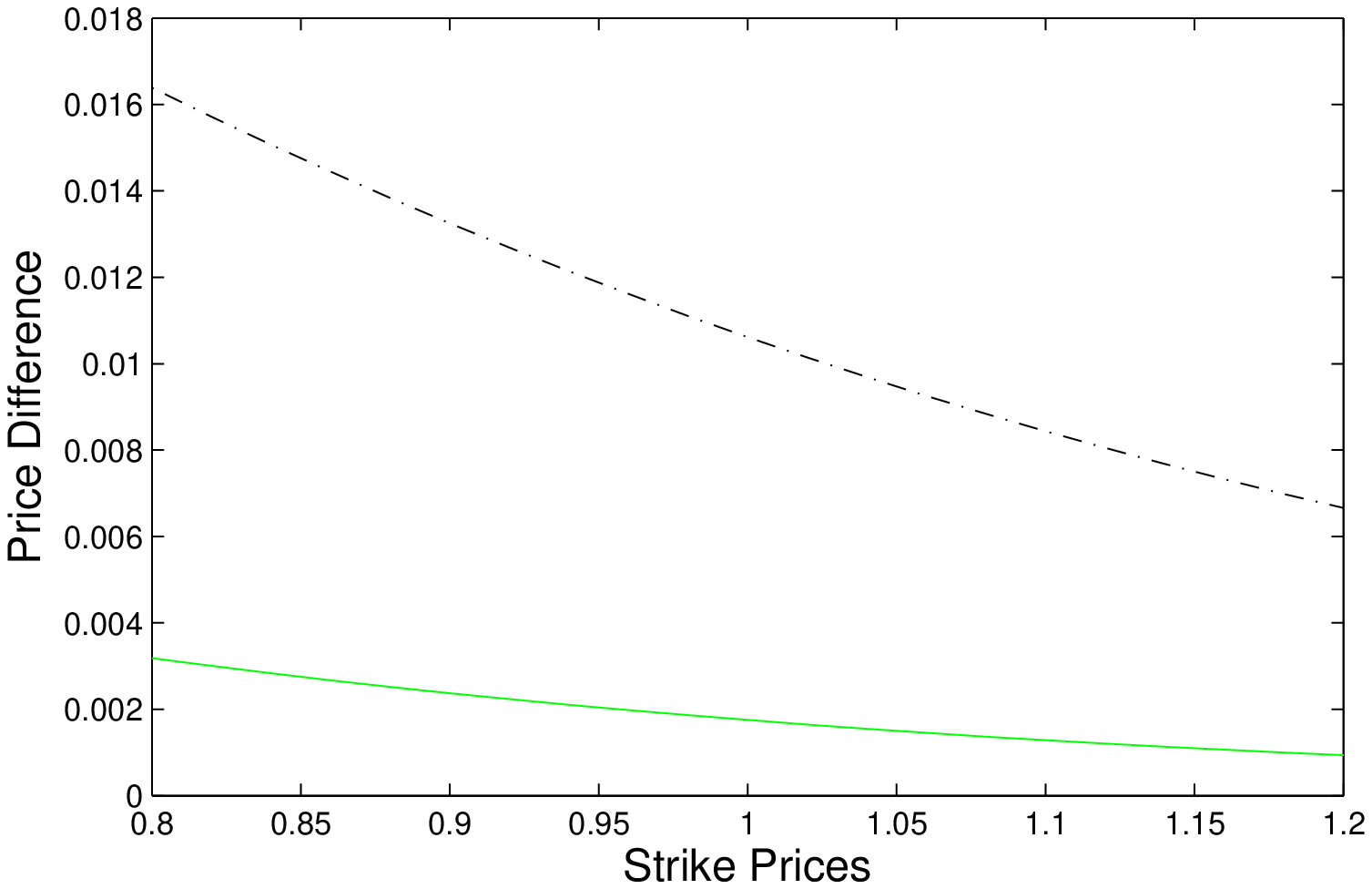}
     \caption{ Option price differences against strike prices.  The solid line corresponds to
     the price difference between the default-free model and  the proposed hybrid model, and the dot-dashed line
     corresponds to the price difference between the default-free model and  the reduced-form model. \label{strikes}}
\end{center}
\end{figure}

\begin{figure}[htbp]
	\centering
	\subfigure[]{
		\includegraphics[width=4.0in]{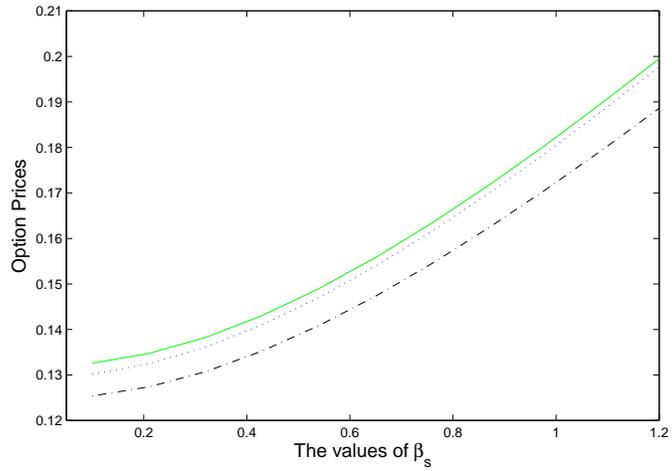}
		%\caption{fig1}
	}
	\quad
	\subfigure[]{
		\includegraphics[width=4.0in]{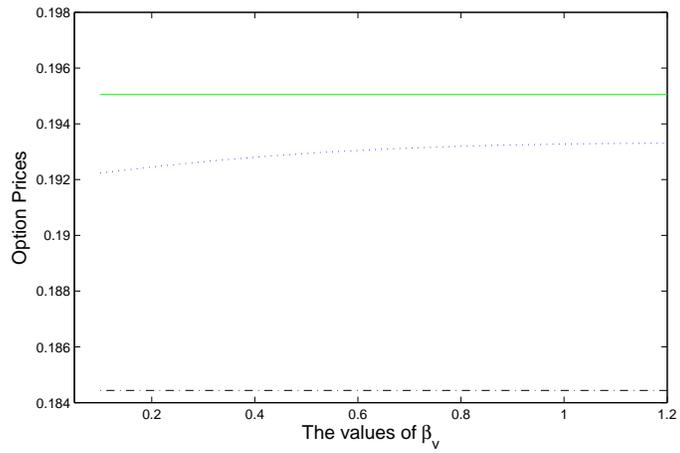}
	}
	\caption{Option prices against the values of $\beta_s$ and $\beta_v$ .  The solid, dotted and dot-dashed lines correspond to
     default-free option prices, option prices in the proposed hybrid model and
     option prices in the reduced-form model, respectively.}\label{betas}
\end{figure}

\begin{figure}[htbp]
	\centering
	\subfigure[]{
		\includegraphics[width=4.0in]{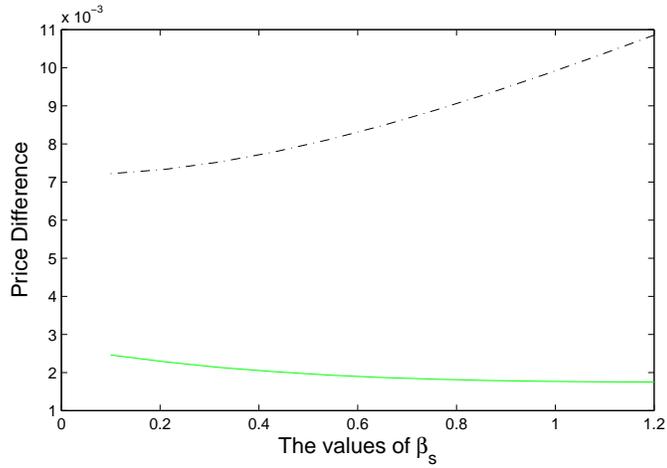}
		%\caption{fig1}
	}
	\quad
	\subfigure[]{
		\includegraphics[width=4.0in]{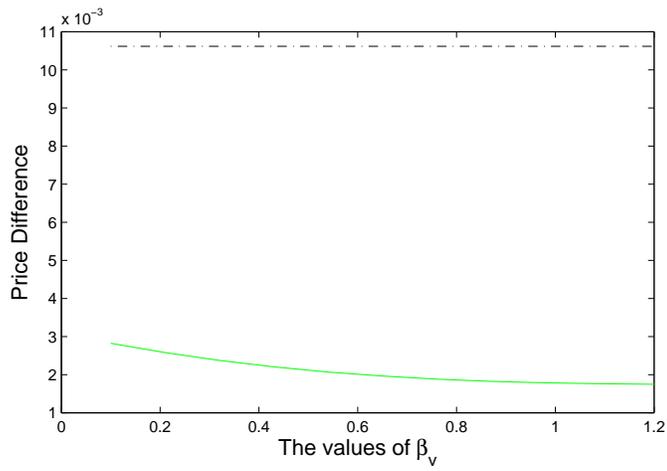}
	}
	\caption{ Option price differences against the values of $\beta_s$ and $\beta_v$.  The solid line corresponds to
     the price difference between the default-free model and  the proposed hybrid model, and the dot-dashed line
     corresponds to the price difference between the default-free model and  the reduced-form model.}\label{diff-betas}
\end{figure}

\begin{figure}
\begin{center}
\makeatletter
\def\@captype{figure}
\makeatother
     \includegraphics[width=4.0in]{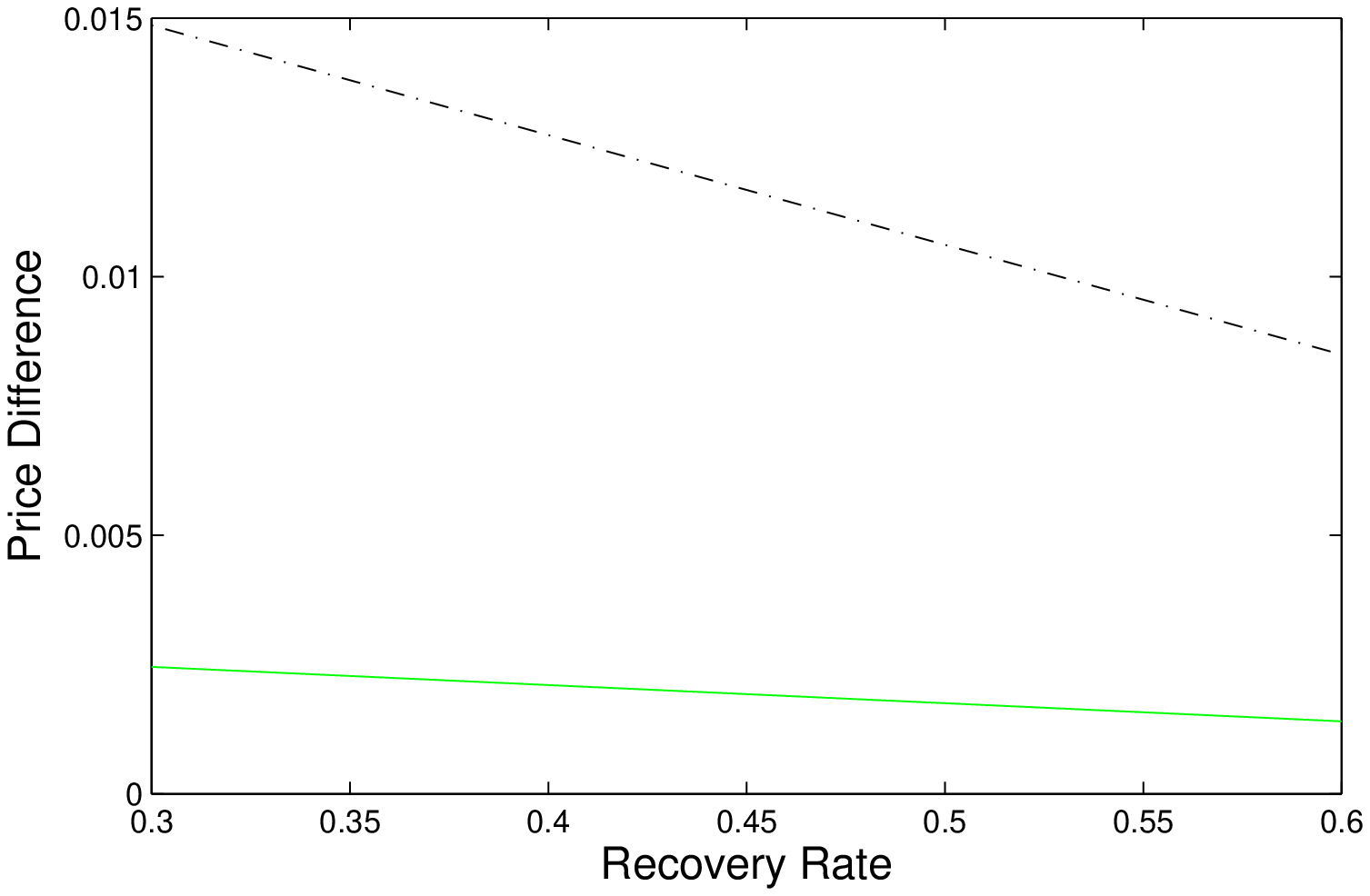}
     \caption{Option price differences against recovery rates.  The solid line corresponds to
     the price difference between the default-free model and  the proposed hybrid model, and the dot-dashed line
     corresponds to the price difference between the default-free model and  the reduced-form model. \label{alpha}}
\end{center}
\end{figure}

\begin{figure}
\begin{center}
\makeatletter
\def\@captype{figure}
\makeatother
     \includegraphics[width=4.0in]{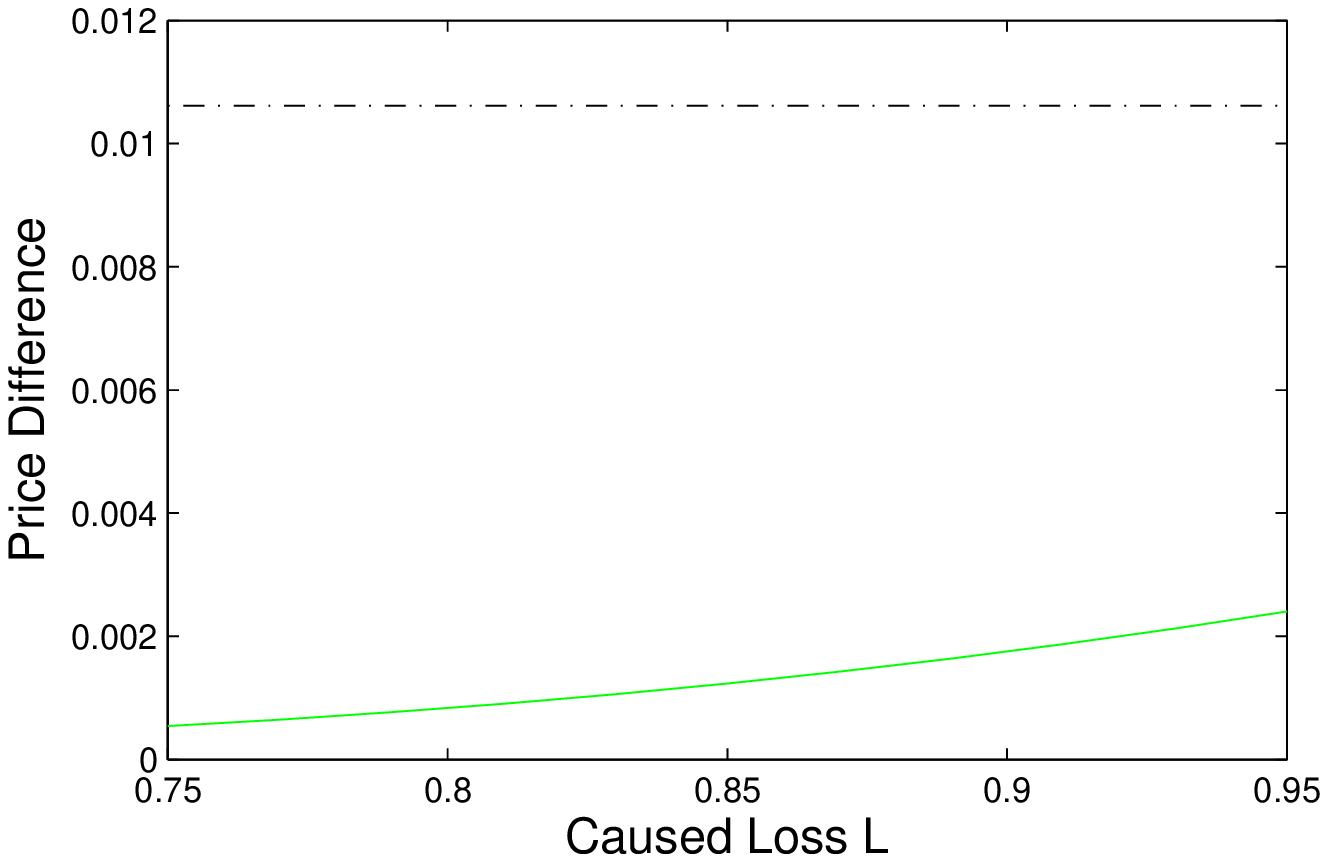}
     \caption{ Option price differences against different caused losses.  The solid line corresponds to
     the price difference between the default-free model and  the proposed hybrid model, and the dot-dashed line
     corresponds to the price difference between the default-free model and  the reduced-form model.\label{DDDD}}
\end{center}
\end{figure}

Figure \ref{T} displays the price difference  with different maturities. The option prices obtained from the proposed hybrid model are close to the option prices without default risk, especially when the maturity is short.
By contrast, default risk in the reduced-form model has a more pronounced effect.
This is because in the reduced-form model default happens when trigger events occur, while
in the hybrid model default happens only  when trigger events occur and the values of the issuer's assets at the arrival time of the trigger events are less than the losses. In other words, default happens more likely in the reduced-form model, thus reducing option prices more significantly.
 Figure \ref{strikes} illustrates  the price difference against different strike prices. A higher strike price will yield a cheaper option. Similar to Figure \ref{T},
 the option has the lowest value in the reduced-form model, as it is more likely to default compared to the hybrid model.

Figure \ref{betas} shows the option values against the sensitivity parameters $\beta_s$ and $\beta_v$, {and the corresponding
price difference is shown in Figure \ref{diff-betas}.}
Recall that $\beta_s$ represents the sensitivity of the stock price against systematic risk.
The option prices will increase with larger $\beta_s$, i.e. with larger systematic risks. Intuitively, with a larger value of $\beta_s$, the value of the underlying asset becomes more volatile. Thus, it is more likely the option is in-the-money and, therefore, its price becomes higher. On the other hand, since $\beta_v$ captures the sensitivity of the issuer's assets to systematic risk, larger $\beta_v$ means the issuer's assets become more risky and, as a result, the issuer is more likely to default. Therefore, one might expect the option prices in the hybrid model become smaller with larger $\beta_v$. However, this is not the case. We observe from Figure \ref{betas}(b) a higher option price with increasing $\beta_v$. This is because larger $\beta_v$ also means the underlying assets and the issuer's assets are more likely to be correlated, which in turn {makes option issuers less likely to default when  call options end in the money, yielding a higher option price consequently.}
%In Figure \ref{betav}, we can observe option prices  with different values of $\beta_v$.
%Although a higher value of $\beta_v$ means that the issuer's asset is more risky, option prices obtained from the proposed hybrid model  increase with
% the values of $\beta_v$. According to the intuition, a more risky asset  of option issuers corresponds to a higher default probability, reducing option prices.
% However, the observed results in  Figure \ref{betav} show that the correlation between the underlying asset and the issuer's asset plays a more important role. A  higher value of $\beta_v$ corresponds to a stronger  correlation,  which enhances the values of the options   obtained from the proposed hybrid model.

Figure \ref{alpha} depicts the price difference  with different recovery rates. Intuitively,  a higher recovery rate corresponds to a higher option price.
However, the effects of recovery rates in the hybrid model are not as significant as those in the reduced-form model.
Figure \ref{DDDD} shows the price difference with different losses (i.e. different values of LGD).
The option prices without default risk and the values of options in the reduced-form model are not affected by the caused losses.
In the hybrid model, it is more likely that default occurs with a higher value of losses, resulting in a lower option price {and a higher default premium.}

\section{Conclusion}

In this paper, we contribute to the literature on  vulnerable options by working under a hybrid credit risk model. The proposed hybrid credit risk model incorporates the features of both structural and reduced-form models.
The dynamics of the market index, as well as the dynamics of the underlying assets and option issuer's assets are driven by Heston-Nandi GARCH processes. The underlying intensity process is
exposed to both systematic risk and idiosyncratic risk.
In this way, all the dynamics are correlated with each other through the systematic risk factor.
Finally, we derive an explicit pricing formula of vulnerable options and
perform numerical analysis to illustrate option prices.

\section*{Acknowledgement}
 The authors would like to thank the anonymous referee and the editor for their helpful comments and valuable suggestions that led to several important improvements. All errors are our responsibility.

\section*{Appendix}
\label{sec:Appendixb} \setcounter{equation}{0}
\renewcommand{\thesection}{A.\arabic{theorem}}
\renewcommand{\thesection}{A.\arabic{lemma}}
\renewcommand{\theequation}{A.\arabic{equation}}
\noindent \textbf{Proof of Proposition \ref{generating function}:}\ \\
We first focus on the case $j\leq t\leq T$. Note that given the information at time $t$, $V(j)$,  $\Lambda(j)$ and $\sum_{k=1}^{j-1} \Lambda(k)$ are all known. Therefore, we obtain that
\begin{eqnarray*}
f(t)&=&E_{t}\Big[\exp\Big\{\phi_1 \ln S(T) +\phi_2\ln V(j) +\phi_3 \Lambda(j) +\phi_4\sum_{k=1}^{j-1} \Lambda(k)  \Big\}\Big]\nonumber\\
&=&e^{\phi_2\ln V(j) +\phi_3 \Lambda(j) +\phi_4\sum_{k=1}^{j-1} \Lambda(k)  }E_{t}\Big[\exp\Big\{\phi_1 \ln S(T) \Big\}\Big].
\end{eqnarray*}
In addition, at time $T$, $S(T)$ is also known, it follows that
\begin{eqnarray*}
f(T)
&=&e^{\phi_2\ln V(j) +\phi_3 \Lambda(j) +\phi_4\sum_{k=1}^{j-1} \Lambda(k)  }E_{T}\Big[\exp\Big\{\phi_1 \ln S(T) \Big\}\Big]\nonumber\\
&=&e^{\phi_1 \ln S(T) +\phi_2\ln V(j) +\phi_3 \Lambda(j) +\phi_4\sum_{k=1}^{j-1} \Lambda(k)  },
\end{eqnarray*}
which in turn implies that
\begin{eqnarray*}
A_0(T)=A_1(T)=A_2(T)=0.
\end{eqnarray*}
According to  the law of iterated expectations, we have that
\begin{eqnarray*}
&&E_{t}\Big[\exp\Big\{\phi_1 \ln S(T) \Big\}\Big]\\
&=& E_t\Big[E_{t+1}\Big[\exp\Big\{\phi_1 \ln S(T) \Big\}\Big]\Big]\\
&=&E_t\Big[\exp\Big\{  \phi_1\ln S(t+1)+A_0(t+1)+A_1(t+1)h_m(t+2)+A_2(t+1)h_s(t+2) \Big\}\Big].
\end{eqnarray*}
Substituting the dynamics of $\ln S(t+1)$, $h_m(t+2)$ and $h_s(t+2)$ yields that
\begin{eqnarray*}
&&E_{t}\Big[\exp\Big\{\phi_1 \ln S(T) \Big\}\Big]\\
&=&E_t\Big[\exp\Big\{  \phi_1\ln S(t+1)+A_0(t+1)+A_1(t+1)h_m(t+2)+A_2(t+1)h_s(t+2) \Big\}\Big]\nonumber\\
&=&E_t\Big[\exp\Big\{ \phi_1\ln S(t)+\phi_1 r-\frac{1}{2}\phi_1 h_s(t+1)+\phi_1\sqrt{h_s(t+1)}Z_s(t+1)\\
&&\ \ \ \ \ \ \ \ \ \ -\frac{1}{2}\phi_1\beta_s^2h_m(t+1)+\phi_1\beta_s\sqrt{h_m(t+1)}Z_m(t+1)+  A_0(t+1)\\
&& +A_1(t+1)\Big(w_m+b_m h_m(t+1)+a_m(Z_m(t+1)-c_m\sqrt{h_m(t+1)})^2\Big)\\
&& + A_2(t+1)\Big(w_s+b_s h_s(t+1)+a_s(Z_s(t+1)-c_s\sqrt{h_s(t+1)})^2\Big)\Big\}\Big].
\end{eqnarray*}
Using the fact that $Ee^{a(Z+b)^2}=e^{-\frac{1}{2}\ln (1-2a)+\frac{ab^2}{1-2a}}$ with $Z$ being a standard normal variable and  some algebra shows that
\begin{align*}
A_0(t)=&  \phi_1 r+A_0(t+1)+ w_m A_1(t+1)+w_sA_2(t+1)-\frac{1}{2}\ln(1-2a_mA_1(t+1))\\
&-\frac{1}{2}\ln(1-2a_sA_2(t+1)),\\
A_1(t)=& b_mA_1(t+1)-\frac{1}{2}\phi_1\beta_s^2
+\phi_1\beta_sc_m-\frac{1}{2}c_m^2+\frac{\frac{1}{2}(\phi_1\beta_s-c_m)^2}{1-2a_mA_1(t+1)},\\
A_2(t)=& b_sA_2(t+1)-\frac{1}{2}\phi_1+\phi_1c_s-\frac{1}{2}c_s^2+\frac{\frac{1}{2}(\phi_1-c_s)^2}{1-2a_sA_2(t+1)}.
\end{align*}
Hence, $A_0(t)$, $A_1(t)$ and $A_2(t)$ ($j\leq t\leq T$) can be obtained recursively with terminal conditions $A_0(T)=A_1(T)=A_2(T)=0$ and the above expressions.

In what follows, we turn to the case $t< j$. Applying  the law of iterated expectations to $f(t)$ yields that
\begin{eqnarray*}
f(t)&=&E_{t}\Big[\exp\Big\{\phi_1 \ln S(T) +\phi_2\ln V(j) +\phi_3 \Lambda(j) +\phi_4\sum_{k=1}^{j-1} \Lambda(k)  \Big\}\Big]\nonumber\\
&=&E_{t}\Big[E_{t+1} \Big[ \exp\Big\{\phi_1 \ln S(T) +\phi_2\ln V(j) +\phi_3 \Lambda(j) +\phi_4\sum_{k=1}^{j-1} \Lambda(k)  \Big\}\Big]\Big]\nonumber\\
&=&E_{t}\Big[f(t+1)\Big]\nonumber\\
&=&E_{t}\Big[\exp\Big\{\phi_2\ln V(t+1) +\phi_4\sum_{k=1}^{t+1} \Lambda(k)  +\phi_1\ln S(t+1)+B_0(t+1)\nonumber\\
 && \ \ \ \ +B_1(t+1)h_m(t+2)+B_2(t+1)h_s(t+2)+B_3(t+1)h_v(t+2)
 + B_4(t+1)\Lambda(t+2)\Big\}\Big].
\end{eqnarray*}
Substituting the dynamics of $\ln V(t+1)$, $\Lambda(t+2)$, $\ln S(t+1)$, $h_m(t+2)$, $h_s(t+2)$ and $h_v(t+2)$ yields that
\begin{eqnarray*}
f(t)
&=&E_{t}\Big[\exp\Big\{\phi_2\ln V(t+1)  +\phi_4\sum_{k=1}^{t+1} \Lambda(k)  +\phi_1\ln S(t+1)+B_0(t+1)\nonumber\\
 && \ \ \ \ +B_1(t+1)h_m(t+2)+B_2(t+1)h_s(t+2)+B_3(t+1)h_v(t+2)+ B_4(t+1)\Lambda(t+2)\Big\}\Big]\\
&=&E_t\Big[\exp\Big\{ \phi_2\ln V(t)+\phi_2 r-\frac{1}{2}\phi_2 h_v(t+1)+\phi_2\sqrt{h_v(t+1)}Z_v(t+1)\\
&&\ \ \ \ \ \ \ \ \ \ -\frac{1}{2}\phi_2\beta_v^2h_m(t+1)+\phi_2\beta_v\sqrt{h_m(t+1)}Z_m(t+1)+\phi_4\sum_{k=1}^{t+1} \Lambda(k)\\
&& \ \ \ \ \ \ \ \ \ \  +\phi_1\ln S(t)+\phi_1 r-\frac{1}{2}\phi_1 h_s(t+1)+\phi_1\sqrt{h_s(t+1)}Z_s(t+1)\\
&&\ \ \ \ \ \ \ \ \ \ -\frac{1}{2}\phi_1\beta_s^2h_m(t+1)+\phi_1\beta_s\sqrt{h_m(t+1)}Z_m(t+1)+  B
_0(t+1)\\
&&\ \ \ \ \ \ \ \ \ \ +B_1(t+1)\Big(w_m+b_m h_m(t+1)+a_m(Z_m(t+1)-c_m\sqrt{h_m(t+1)})^2\Big)\\
&& \ \ \ \ \ \ \ \ \ \ + B_2(t+1)\Big(w_s+b_s h_s(t+1)+a_s(Z_s(t+1)-c_s\sqrt{h_s(t+1)})^2\Big)\\
&& \ \ \ \ \ \ \ \ \ \ +B_3(t+1)\Big(w_v+b_v h_v(t+1)+a_v(Z_v(t+1)-c_v\sqrt{h_v(t+1)})^2\Big)\\
&&\ \ \ \ \ \ \ \ \ \ + B_4(t+1)\Big(w_{\lambda}+b_{\lambda} \Lambda(t+1)+a_\lambda(Z_{m}(t+1))^2+    c_\lambda(Z_{v}(t+1))^2\Big)\Big\}\Big].
\end{eqnarray*}
Rearranging terms implies that
\begin{eqnarray*}
f(t)
&=&E_t\Big[\exp\Big\{ \phi_2\ln V(t)+\phi_4\sum_{k=1}^{t} \Lambda(k)    +\phi_1\ln S(t)+  B
_0(t+1)+   (\phi_2+\phi_1) r\\
&&\ \ \ \ \ \ \ \ \ \ +w_m B_1(t+1)+
w_sB_2(t+1)+w_vB_3(t+1)+w_{\lambda}B_4(t+1)\\
&&\ \ \ \ \ \ \ \ \ \ +(b_mB_1(t+1)-\frac{1}{2}\phi_2\beta_v^2-\frac{1}{2}\phi_1\beta_s^2) h_m(t+1)\\
&&\ \ \ \ \ \ \ \ \ \  +(b_vB_3(t+1)-\frac{1}{2}\phi_2) h_v(t+1)+(b_sB_2(t+1)-\frac{1}{2}\phi_1) h_s(t+1)\\
&&\ \ \ \ \ \ \ \ \ \ + (b_{\lambda} B_4(t+1)+\phi_4)\Lambda(t+1) +\Phi_m+\Phi_s+\Phi_v\Big\}\Big],
\end{eqnarray*}
where
\begin{eqnarray*}
\Phi_s&=&\phi_1\sqrt{h_s(t+1)}Z_s(t+1)+a_sB_2(t+1)(Z_s(t+1)-c_s\sqrt{h_s(t+1)})^2,\\
\Phi_v&=&\phi_2\sqrt{h_v(t+1)}Z_v(t+1)+a_v B_3(t+1)(Z_v(t+1)-c_v\sqrt{h_v(t+1)})^2\\
&&\ \ \ \ \ \ \ \ \ \ +c_\lambda B_4(t+1) (Z_{v}(t+1))^2,\\
\Phi_m&=&(\phi_2\beta_v+ \phi_1\beta_s )\sqrt{h_m(t+1)}Z_m(t+1)+ a_mB_1(t+1)(Z_m(t+1)-c_m\sqrt{h_m(t+1)})^2 \\
&&\ \ \ \ \ \ \ \ \ \ + a_\lambda B_4(t+1)(Z_{m}(t+1))^2.
\end{eqnarray*}
In order to obtain the explicit expression of $f(t)$, we only need to calculate $E_t[e^{\Phi_m+\Phi_s+\Phi_v}]=E_t[e^{\Phi_m}]E_t[e^{\Phi_s}]E_t[e^{\Phi_v}] $. Note that $\Phi_s$, $\Phi_m$ and $\Phi_v$ have similar forms and all can be obtained based on  the following form,
\begin{eqnarray*}
E[\exp\{    \mu_1 \sqrt{h}Z +\mu_2(Z-\mu_3\sqrt{h})^2+\mu_4Z^2\}],
\end{eqnarray*}
where $\mu_1$, $\mu_2$, $\mu_3$ and $\mu_4$ are all constants and $Z$ is a standard normal variable. Using the fact that  $Ee^{a(Z+b)^2}=e^{-\frac{1}{2}\ln (1-2a)+\frac{ab^2}{1-2a}}$, we have that
\begin{eqnarray}\label{expectation}
&&E[\exp\{   \mu_1 \sqrt{h}Z +\mu_2(Z-\mu_3\sqrt{h})^2+\mu_4Z^2\}]\nonumber\\
&=& E[\exp\{   ( \mu_2 +\mu_4)Z^2-2(\mu_2\mu_3-\mu_1/2 ) Z\sqrt{h}+\mu_2\mu_3^2 h     \}]\nonumber\\
&=&E[\exp\{   ( \mu_2 +\mu_4)\Big( Z- \frac{\mu_2\mu_3-\mu_1/2 }{\mu_2 +\mu_4}\sqrt{h}\Big)^2-\frac{(\mu_2\mu_3-\mu_1/2 )^2}{\mu_2 +\mu_4}h+\mu_2\mu_3^2 h \}]\nonumber\\
&=&e^{\mu_2\mu_3^2h-\frac{(\mu_2\mu_3-\mu_1/2 )^2}{\mu_2 +\mu_4}h} E[\exp\{   ( \mu_2 +\mu_4)\Big( Z- \frac{\mu_2\mu_3-\mu_1/2 }{\mu_2 +\mu_4}\sqrt{h}\Big)^2 \}]\nonumber\\
&=&\exp\{ \mu_2\mu_3^2h-\frac{(\mu_2\mu_3-\mu_1/2 )^2}{\mu_2 +\mu_4}h-\frac{1}{2}\ln (1-2( \mu_2 +\mu_4 ))   +\frac{( \mu_2 +\mu_4 )(\frac{\mu_2\mu_3-\mu_1/2 }{\mu_2 +\mu_4})^2}{1-2( \mu_2 +\mu_4 )}h \}\nonumber\\
&=&\exp\{ -\frac{1}{2}\ln (1-2( \mu_2 +\mu_4 )) +\Big(\mu_2\mu_3^2-\frac{(\mu_2\mu_3-\mu_1/2 )^2}{\mu_2 +\mu_4}  +\frac{( \mu_2 +\mu_4 )(\frac{\mu_2\mu_3-\mu_1/2 }{\mu_2 +\mu_4})^2}{1-2( \mu_2 +\mu_4 )}\Big)h \}\nonumber\\
&=&\exp\{ -\frac{1}{2}\ln (1-2( \mu_2 +\mu_4 )) +\Big(\mu_2\mu_3^2 +\frac{2(\mu_2\mu_3-\mu_1/2 )^2}{1-2( \mu_2 +\mu_4 )}\Big)h \}.
\end{eqnarray}
Therefore, we can write $f(t)$ in the following form
\begin{eqnarray*}
f(t)
&=&\exp\Big\{\phi_2\ln V(t)+\phi_4\sum_{k=1}^{t} \Lambda(k)  +\phi_1\ln S(t)+B_0(t)\nonumber\\
 && \ \ \ \ \ \ \ \ \ \ \ +B_1(t)h_m(t+1)+B_2(t)h_s(t+1)+B_3(t)h_v(t+1)+B_4(t+1)  \Lambda(t+1) \Big\},
\end{eqnarray*}
where
\begin{eqnarray*}
B_0(t)&=&B_0(t+1)+   (\phi_2+\phi_1) r +w_m B_1(t+1)+w_sB_2(t+1)+w_vB_3(t+1)+w_{\lambda}B_4(t+1)\\
&&\ \ \ -\frac{1}{2}\ln (1-2 (a_mB_1(t+1)+a_\lambda B_4(t+1) ))-\frac{1}{2}\ln (1-2 a_sB_2(t+1)))\\
&&\ \ \  -\frac{1}{2}\ln (1-2 (a_vB_3(t+1)+c_\lambda B_4(t+1) )),\\
B_1(t)&=&b_mB_1(t+1)-\frac{1}{2}\phi_2\beta_v^2-\frac{1}{2}\phi_1\beta_s^2+ a_mc_m^2B_1(t+1)\\
&&\ \ \ +\frac{2(a_mc_mB_1(t+1)-(\phi_2\beta_v+ \phi_1\beta_s )/2)^2 }{1-2(a_mB_1(t+1)+a_\lambda B_4(t+1) )},\\
B_2(t)&=&b_sB_2(t+1)-\frac{1}{2}\phi_1 + a_sc_s^2B_2(t+1)+\frac{2(a_sc_sB_2(t+1)- \phi_1/2)^2 }{1-2a_sB_2(t+1)},\\
B_3(t)&=&b_vB_3(t+1)-\frac{1}{2}\phi_2+ a_vc_v^2B_3(t+1)+\frac{2(a_vc_vB_3(t+1)- \phi_2/2)^2 }{1-2(a_vB_3(t+1)+c_\lambda B_4(t+1) )},\\
B_4(t)&=& b_{\lambda}B_4(t+1)+\phi_4.
\end{eqnarray*}
Now we need the terminal conditions of $B_{k}(t),\ k=0,1,2,3,4$ ($t< j$). In other words, we need to determine the values of $B_{k}(j-1),\ k=0,1,2,3,4$.
 Actually, we already have the expression of $f(j)$ from the case $j\leq t\leq T$ we previously considered,
\begin{eqnarray*}
f(j)
&=&\exp\Big\{\phi_2\ln V(j) +\phi_3 \Lambda(j) +\phi_4\sum_{k=1}^{j-1} \Lambda(k)  +\phi_1\ln S(j)+A_0(j)\nonumber\\
 && \ \ \ \ \ \ \ \ \ \ \ +A_1(j)h_m(j+1)+A_2(j)h_s(j+1)\Big\}.
\end{eqnarray*}
According to  the law of iterated expectations, we have that
\begin{eqnarray*}
f(j-1)&=&E_{j-1}\Big[f(j)\Big]\\
&=&E_{j-1}\Big[\exp\Big\{\phi_2\ln V(j) +\phi_3 \Lambda(j) +\phi_4\sum_{k=1}^{j-1} \Lambda(k)  +\phi_1\ln S(j)+A_0(j)\nonumber\\
 && \ \ \ \ \ \ \ \ \ \ \ +A_1(j)h_m(j+1)+A_2(j)h_s(j+1)\Big\}\Big].
\end{eqnarray*}
Substituting the dynamics of $\ln V(j)$, $\ln S(j)$, $h_m(j+1)$, and $h_s(j+1)$ and using (\ref{expectation}) imply that
\begin{eqnarray*}
f(j-1)
&=&E_{j-1}\Big[\exp\Big\{\phi_2\ln V(j) +\phi_3 \Lambda(j) +\phi_4\sum_{k=1}^{j-1} \Lambda(k)  +\phi_1\ln S(j)+A_0(j)\nonumber\\
 && \ \ \ \ \ \ \ \ \ \ \ +A_1(j)h_m(j+1)+A_2(j)h_s(j+1)\Big\}\Big]\nonumber\\
&=&E_t\Big[\exp\Big\{ \phi_2\ln V(j-1)+\phi_2 r-\frac{1}{2}\phi_2 h_v(j)+\phi_2\sqrt{h_v(j)}Z_v(j)\\
&&\ \ \ \ \ \ \ \ \ \ -\frac{1}{2}\phi_2\beta_v^2h_m(j)+\phi_2\beta_v\sqrt{h_m(j)}Z_m(j)+\phi_3 \Lambda(j)+\phi_4\sum_{k=1}^{j-1} \Lambda(k)\\
&& \ \ \ \ \ \ \ \ \ \  +\phi_1\ln S(j-1)+\phi_1 r-\frac{1}{2}\phi_1 h_s(j)+\phi_1\sqrt{h_s(j)}Z_s(j)\\
&&\ \ \ \ \ \ \ \ \ \ -\frac{1}{2}\phi_1\beta_s^2h_m(j)+\phi_1\beta_s\sqrt{h_m(j)}Z_m(j)+ A_0(j)\\
&&\ \ \ \ \ \ \ \ \ \ +A_1(j)\Big(w_m+b_m h_m(j)+a_m(Z_m(j)-c_m\sqrt{h_m(j)})^2\Big)\\
&& \ \ \ \ \ \ \ \ \ \ + A_2(j)\Big(w_s+b_s h_s(j)+a_s(Z_s(j)-c_s\sqrt{h_s(j)})^2\Big)\Big\}\Big]\\
&=&\exp\Big\{\phi_2\ln V(j-1)+\phi_4\sum_{k=1}^{j-1} \Lambda(k)  +\phi_1\ln S(j-1)+B_0(j-1)\nonumber\\
 && \ \ \ \ \ \ \ \ \ \ \ +B_1(j-1)h_m(j)+B_2(j-1)h_s(j)+B_3(j-1)h_v(j)+B_4(j-1)  \Lambda(j) \Big\},
\end{eqnarray*}
where
\begin{eqnarray*}
B_0(j-1)&=&A_0(j)+   (\phi_2+\phi_1) r +w_m A_1(j)+w_sA_2(j)\\
&&-\frac{1}{2}\ln (1-2 a_mA_1(j))-\frac{1}{2}\ln (1-2 a_sA_2(j)),\\
B_1(j-1)&=&b_mA_1(j)-\frac{1}{2}\phi_2\beta_v^2-\frac{1}{2}\phi_1\beta_s^2+ a_mc_m^2A_1(j) +\frac{2(a_mc_mA_1(j)-(\phi_2\beta_v+ \phi_1\beta_s )/2)^2 }{1-2a_mA_1(j)},\\
B_2(j-1)&=&b_sA_2(j)-\frac{1}{2}\phi_1 + a_sc_s^2A_2(j)+\frac{2(a_sc_sA_2(j)- \phi_1/2)^2 }{1-2a_sA_2(j)},\\
B_3(j-1)&=&-\frac{1}{2}\phi_2+\frac{1}{2}\phi_2^2,\\
B_4(j-1)&=& \phi_3.
\end{eqnarray*}
This completes the proof of the proposition.
\hfill$\Box$\\

\noindent \textbf{Proof of Theorem \ref{option prices}:}\ \\
First, we deal with the term $E\Big[(S(T)-K)^+\Big]$. Recall the definition of $f(t;\phi_1,\phi_2,\phi_3,\phi_4)$ and note that $f(0;i\phi_1,0,0,0)$ is the characteristic function of $\ln S(T)$ under $Q$.
From standard probability theory (see, e.g., Kendall and Stuart (1977)), we can obtain  the distribution function of $\ln S(T)$, that is,
$$Q(\ln S(T)\leq x)=\frac{1}{2}-\frac{1}{\pi}\int_{0}^\infty \textrm{Re}\Big[\frac{e^{-i \phi_1 x}f(0;i\phi_1,0,0,0)}{ i\phi_1 }\Big]\D \phi_1,$$
which in turn implies that
\begin{eqnarray}\label{Q S}
Q(\ln S(T)\geq x)&=&1- Q(\ln S(T)\leq x)\nonumber\\
&=&\frac{1}{2}+\frac{1}{\pi}\int_{0}^\infty \textrm{Re}\Big[\frac{e^{-i \phi_1 x}f(0;i\phi_1,0,0,0)}{ i\phi_1 }\Big]\D \phi_1.
\end{eqnarray}
 The term  $E\Big[(S(T)-K)^+\Big]$  can be derived  after introducing  a new probability measure $Q_1$ defined by
\begin{eqnarray*}
Q_1(O)=\frac{E\Big[I(O)S(T)\Big]}{E\Big[S(T)\Big]},
\end{eqnarray*}
for any event $O\in \mathscr{F}_{T}$. Obviously, the characteristic function of $\ln S(T)$ under $Q_1$ is given  by
\begin{eqnarray*}
E^{Q_1}\Big[e^{i\phi_1 \ln S(T)}\Big]%&=&E\Big[\frac{S(T)}{E\Big[S(T)\Big]}e^{i\phi_1 \ln S(T)}\Big]\nonumber\\
%&=&\frac{1}{E\Big[S(T)\Big]}E\Big[e^{(1+i\phi_1) \ln S(T)}\Big]\nonumber\\
&=&\frac{f(0;1+i\phi_1,0,0,0)}{f(0;1,0,0,0)}.
\end{eqnarray*}
In addition,  under $Q_1$, it holds that
\begin{eqnarray}\label{Q1 S}
Q_1(\ln S(T)\geq x)
&=&\frac{1}{2}+\frac{1}{\pi}\int_{0}^\infty \textrm{Re}\Big[\frac{e^{-i \phi_1 x}f(0;1+i\phi_1,0,0,0)/f(0;1,0,0,0)}{ i\phi_1 }\Big]\D \phi_1.
\end{eqnarray}
Hence, (\ref{Q S}) and (\ref{Q1 S}) imply that
\begin{eqnarray}\label{term 1}
E\Big[(S(T)-K)^+\Big]&=& E\Big[(S(T)-K)^+\Big]\nonumber\\
&=& E\Big[(S(T)-K)I(\ln S(T)\geq \ln K)\Big]\nonumber\\
&=&   E\Big[S(T)I(\ln S(T)\geq \ln K)\Big]    -K   E\Big[I(\ln S(T)\geq \ln K)\Big]     \nonumber\\
&=&  E[S(T)] E^{Q_1}\Big[I(\ln S(T)\geq \ln K)\Big]    -K   E\Big[I(\ln S(T)\geq \ln K)\Big]     \nonumber\\
&=&  E[S(T)]Q_1(\ln S(T)\geq \ln K)    -K   Q(\ln S(T)\geq \ln K)    \Big) \nonumber\\
&=&  \frac{1}{2}f(0;1,0,0,0)+\frac{1}{\pi}\int_{0}^\infty \textrm{Re}\Big[\frac{e^{-i \phi_1 \ln K}f(0;1+i\phi_1,0,0,0)}{ i\phi_1 }\Big]\D \phi_1\nonumber\\
&&\ \ \ \ \ \ \ \ \ \ \ \ \     - \frac{K}{2}-\frac{K}{\pi}\int_{0}^\infty \textrm{Re}\Big[\frac{e^{-i \phi_1 \ln K}f(0;i\phi_1,0,0,0)}{ i\phi_1 }\Big]\D \phi_1,
\end{eqnarray}
where in the last equality we used (\ref{Q S}) and (\ref{Q1 S}).

Next, we focus on the term  $E\Big[e^{-\sum_{k=1}^{j} \Lambda(k)}I( V(j)<L)  (S(T)-K)^+ \Big]$. We rewrite it as follows:
\begin{eqnarray*}
&&E\Big[e^{-\sum_{k=1}^{j} \Lambda(k)}I( V(j)<L)  (S(T)-K)^+ \Big]\\
&=&   E\Big[e^{-\sum_{k=1}^{j} \Lambda(k)+\ln S(T)}I( V(j)<L, \ln S(T)\geq \ln K)\Big] \nonumber\\
   && \ \ \ -K   E\Big[e^{-\sum_{k=1}^{j} \Lambda(k)}I( V(j)<L, \ln S(T)\geq \ln K)\Big]\nonumber\\
   &=&   E\Big[e^{-\sum_{k=1}^{j} \Lambda(k)+\ln S(T)}I(-\ln V(j)> -\ln L, \ln S(T)\geq \ln K)\Big] \nonumber\\
   && \ \ \ -K   E\Big[e^{-\sum_{k=1}^{j} \Lambda(k)}I( -\ln V(j)> -\ln L, \ln S(T)\geq \ln K)\Big].
\end{eqnarray*}
In the following, we deal with the two parts in the above equality separately. To this end, we define a new probability measure
\begin{eqnarray*}
Q_2(O)=\frac{E\Big[I(O)e^{-\sum_{k=1}^{j} \Lambda(k)+\ln S(T)}\Big]}{E\Big[e^{-\sum_{k=1}^{j} \Lambda(k)+\ln S(T)}\Big]},
\end{eqnarray*}
for any event $O\in \mathscr{F}_{T}$. Under $Q_2$, we have
the joint  characteristic function of $-\ln V(j)$ and $\ln S(T)$ as follows:
\begin{eqnarray*}
E^{Q_2}\Big[e^{i\phi_2(-\ln V(j))+ i\phi_1\ln S(T)}\Big]&=& E\Big[   \frac{e^{-\sum_{k=1}^{j} \Lambda(k)+\ln S(T)}}{E\Big[e^{-\sum_{k=1}^{j} \Lambda(k)+\ln S(T)}\Big]}        e^{i\phi_2(-\ln V(j))+ i\phi_1\ln S(T)}\Big]\nonumber\\
&=& \frac{1}{E\Big[e^{-\sum_{k=1}^{j} \Lambda(k)+\ln S(T)}\Big]}  E\Big[e^{(i\phi_1+1)\ln S(T) -i\phi_2\ln V(j)-\sum_{k=1}^{j} \Lambda(k) }\Big]\nonumber\\
&=&\frac{f(0;i\phi_1+1,-i\phi_2,-1,-1)}{f(0;1,0,-1,-1)}.
\end{eqnarray*}
By inverting the characteristic function, we have that
\begin{eqnarray}\label{pi1}
 \Pi_{j,1}&:=&E\Big[e^{-\sum_{k=1}^{j} \Lambda(k)+\ln S(T)}I(-\ln V(j)> -\ln L, \ln S(T)\geq \ln K)\Big]\nonumber\\
 &=&E\Big[e^{-\sum_{k=1}^{j} \Lambda(k)+\ln S(T)}\Big]E^{Q_2}\Big[I(-\ln V(j)>-\ln L, \ln S(T)\geq \ln K)\Big]\nonumber\\
  &=&f(0;1,0,-1,-1)Q_2\Big(-\ln V(j)> -\ln L, \ln S(T)\geq \ln K\Big)\nonumber\\
 &=& \frac{1}{4}f(0;1,0,-1,-1)+\frac{1}{2\pi}\int_{0}^\infty \textrm{Re}\Big[\frac{e^{-i \phi_1 \ln K}f(0;i\phi_1+1,0,-1,-1)}{ i\phi_1 }\Big]\D \phi_1\nonumber\\
&&\ +\frac{1}{2\pi}\int_{0}^\infty \textrm{Re}\Big[\frac{e^{i \phi_2 \ln L}f(0;1,-i\phi_2,-1,-1)}{ i\phi_2 }\Big]\D \phi_2\nonumber\\
&&\ -\frac{1}{2\pi^2}\int_{0}^\infty\int_0^\infty\Big(\textrm{Re}\Big[\frac{e^{-i \phi_1 \ln K+i  \phi_2 \ln L}f(0;i\phi_1+1,-i\phi_2,-1,-1)}{ \phi_1 \phi_2}\Big]\nonumber\\
&&\ \ \ \ \ \ \ \ \ \ \ \ \ \ \ \ \ \ \ \ \ \ \ -\textrm{Re}\Big[\frac{e^{-i \phi_1 \ln K-i  \phi_2 \ln L}f(0;i\phi_1+1,i\phi_2,-1,-1)}{ \phi_1 \phi_2}\Big]\Big)\D \phi_1\D \phi_2.
\end{eqnarray}
Likewise, we work under $\bar{Q}_2$ defined by
\begin{eqnarray*}
\bar{Q}_2(O)=\frac{E\Big[I(O)e^{-\sum_{k=1}^{j} \Lambda(k)+\ln S(T)}\Big]}{E\Big[e^{-\sum_{k=1}^{j} \Lambda(k)+\ln S(T)}\Big]},
\end{eqnarray*}
for any event $O\in \mathscr{F}_{T}$, and obtain that
\begin{eqnarray}\label{pi2}
 \Pi_{j,2}&:=&E\Big[e^{-\sum_{k=1}^{j} \Lambda(k)}I(-\ln V(j)> -\ln L, \ln S(T)\geq \ln K)\Big]\nonumber\\
   &=&f(0;0,0,-1,-1)\bar{Q}_2\Big(-\ln V(j)> -\ln L, \ln S(T)\geq \ln K\Big)\nonumber\\
   &=& \frac{1}{4}f(0;0,0,-1,-1)+\frac{1}{2\pi}\int_{0}^\infty \textrm{Re}\Big[\frac{e^{-i \phi_1 \ln K}f(0;i\phi_1,0,-1,-1)}{ i\phi_1 }\Big]\D \phi_1\nonumber\\
&&\ +\frac{1}{2\pi}\int_{0}^\infty \textrm{Re}\Big[\frac{e^{i \phi_2 \ln L}f(0;0,-i\phi_2,-1,-1)}{ i\phi_2 }\Big]\D \phi_2\nonumber\\
&&\ -\frac{1}{2\pi^2}\int_{0}^\infty\int_0^\infty\Big(\textrm{Re}\Big[\frac{e^{-i \phi_1 \ln K+i  \phi_2 \ln L}f(0;i\phi_1,-i\phi_2,-1,-1)}{ \phi_1 \phi_2}\Big]\nonumber\\
&&\ \ \ \ \ \ \ \ \ \ \ \ \ \ \ \ \ \ \ \ \ \ \ -\textrm{Re}\Big[\frac{e^{-i \phi_1 \ln K-i  \phi_2 \ln L}f(0;i\phi_1,i\phi_2,-1,-1)}{ \phi_1 \phi_2}\Big]\Big)\D \phi_1\D \phi_2.
\end{eqnarray}
Hence,  it holds that
\begin{eqnarray*}\label{term 2}
&&E\Big[e^{-\sum_{k=1}^{j} \Lambda(k)}I( V(j)<L)  (S(T)-K)^+ \Big]\\
   &=&   E\Big[e^{-\sum_{k=1}^{j} \Lambda(k)+\ln S(T)}I(-\ln V(j)> -\ln L, \ln S(T)\geq \ln K)\Big] \nonumber\\
   && \ \ \ -K   E\Big[e^{-\sum_{k=1}^{j} \Lambda(k)}I( -\ln V(j)> -\ln L, \ln S(T)\geq \ln K)\Big]\nonumber\\
     &=& \Pi_{j,1} -K\Pi_{j,2}.
\end{eqnarray*}
Similarly,
\begin{eqnarray*}\label{term 3}
&&E\Big[e^{-\sum_{k=1}^{j-1} \Lambda(k)}I( V(j)<L)  (S(T)-K)^+ \Big]\\
   &=&   E\Big[e^{-\sum_{k=1}^{j-1} \Lambda(k)+\ln S(T)}I(-\ln V(j)> -\ln L, \ln S(T)\geq \ln K)\Big] \nonumber\\
   && \ \ \ -K   E\Big[e^{-\sum_{k=1}^{j-1} \Lambda(k)}I( -\ln V(j)> -\ln L, \ln S(T)\geq \ln K)\Big]\nonumber\\
    &=& \Pi_{j,3} -K\Pi_{j,4},
\end{eqnarray*}
where
\begin{eqnarray}\label{pi3}
 \Pi_{j,3}&:=&\frac{1}{4}f(0;1,0,0,-1)+\frac{1}{2\pi}\int_{0}^\infty \textrm{Re}\Big[\frac{e^{-i \phi_1 \ln K}f(0;i\phi_1+1,0,0,-1)}{ i\phi_1 }\Big]\D \phi_1\nonumber\\
&&\ +\frac{1}{2\pi}\int_{0}^\infty \textrm{Re}\Big[\frac{e^{i \phi_2 \ln L}f(0;1,-i\phi_2,0,-1)}{ i\phi_2 }\Big]\D \phi_2\nonumber\\
&&\ -\frac{1}{2\pi^2}\int_{0}^\infty\int_0^\infty\Big(\textrm{Re}\Big[\frac{e^{-i \phi_1 \ln K+i  \phi_2 \ln L}f(0;i\phi_1+1,-i\phi_2,0,-1)}{ \phi_1 \phi_2}\Big]\nonumber\\
&&\ \ \ \ \ \ \ \ \ \ \ \ \ \ \ \ \ \ \ \ \ \ \ -\textrm{Re}\Big[\frac{e^{-i \phi_1 \ln K-i  \phi_2 \ln L}f(0;i\phi_1+1,i\phi_2,0,-1)}{ \phi_1 \phi_2}\Big]\Big)\D \phi_1\D \phi_2,
\end{eqnarray}
and
\begin{eqnarray}\label{pi4}
 \Pi_{j,4}&:=& \frac{1}{4}f(0;0,0,0,-1)+\frac{1}{2\pi}\int_{0}^\infty \textrm{Re}\Big[\frac{e^{-i \phi_1 \ln K}f(0;i\phi_1,0,0,-1)}{ i\phi_1 }\Big]\D \phi_1\nonumber\\
&&\ +\frac{1}{2\pi}\int_{0}^\infty \textrm{Re}\Big[\frac{e^{i \phi_2 \ln L}f(0;0,-i\phi_2,0,-1)}{ i\phi_2 }\Big]\D \phi_2\nonumber\\
&&\ -\frac{1}{2\pi^2}\int_{0}^\infty\int_0^\infty\Big(\textrm{Re}\Big[\frac{e^{-i \phi_1 \ln K+i  \phi_2 \ln L}f(0;i\phi_1,-i\phi_2,0,-1)}{ \phi_1 \phi_2}\Big]\nonumber\\
&&\ \ \ \ \ \ \ \ \ \ \ \ \ \ \ \ \ \ \ \ \ \ \ -\textrm{Re}\Big[\frac{e^{-i \phi_1 \ln K-i  \phi_2 \ln L}f(0;i\phi_1,i\phi_2,0,-1)}{ \phi_1 \phi_2}\Big]\Big)\D \phi_1\D \phi_2.
\end{eqnarray}
Note that $\Pi_{j,3}$ and $\Pi_{j,4}$ have similar forms as $\Pi_{j,1}$ and $\Pi_{j,2}$, and they can be obtained by
replacing $f(0;\cdot,\cdot,-1,\cdot)$  in  $\Pi_{j,1}$ and $\Pi_{j,2}$  with $f(0;\cdot,\cdot,0,\cdot)$, respectively.

We next calculate $E\Big[e^{-\sum_{k=1}^{j} \Lambda(k)}I( V(j)<L) V(j)  (S(T)-K)^+ \Big]$. Note that $E\Big[e^{-\sum_{k=1}^{j-1} \Lambda(k)}I( V(j)<L) V(j)  (S(T)-K)^+ \Big]$ can  be calculated in a similar way. To this end, define another probability measure $Q_3$ as follows:
\begin{eqnarray*}
Q_3(O)=\frac{E\Big[I(O)e^{-\sum_{k=1}^{j} \Lambda(k)+\ln S(T)+\ln V(j)}\Big]}{E\Big[e^{-\sum_{k=1}^{j} \Lambda(k)+\ln S(T)+\ln V(j)}\Big]},
\end{eqnarray*}
for any event $O\in \mathscr{F}_{T}$.
The joint  characteristic function of $-\ln V(j)$ and $\ln S(T)$ under $Q_3$ is
\begin{eqnarray*}
E^{Q_3}\Big[e^{i\phi_2(-\ln V(j))+ i\phi_1\ln S(T)}\Big]&=& E\Big[   \frac{e^{-\sum_{k=1}^{j} \Lambda(k)+\ln S(T)+\ln V(j)}}{E\Big[e^{-\sum_{k=1}^{j} \Lambda(k)+\ln S(T)+\ln V(j)}\Big]}        e^{i\phi_2(-\ln V(j))+ i\phi_1\ln S(T)}\Big]\nonumber\\
&=&\frac{f(0;i\phi_1+1,-i\phi_2+1,-1,-1)}{f(0;1,1,-1,-1)}.
\end{eqnarray*}
By inverting the characteristic function, we obtain that
\begin{eqnarray}\label{pi5}
 \Pi_{j,5}&:=&E\Big[e^{-\sum_{k=1}^{j} \Lambda(k)+\ln S(T)}V(j)I(-\ln V(j)> -\ln L, \ln S(T)\geq \ln K)\Big]\nonumber\\
 &=&E\Big[e^{-\sum_{k=1}^{j} \Lambda(k)+\ln S(T)+\ln V(j)}\Big]E^{Q_3}\Big[I(-\ln V(j)>-\ln L, \ln S(T)\geq \ln K)\Big]\nonumber\\
  &=& \frac{1}{4}f(0;1,1,-1,-1)+\frac{1}{2\pi}\int_{0}^\infty \textrm{Re}\Big[\frac{e^{-i \phi_1 \ln K}f(0;i\phi_1+1,1,-1,-1)}{ i\phi_1 }\Big]\D \phi_1\nonumber\\
&&\ +\frac{1}{2\pi}\int_{0}^\infty \textrm{Re}\Big[\frac{e^{i \phi_2 \ln L}f(0;1,-i\phi_2+1,-1,-1)}{ i\phi_2 }\Big]\D \phi_2\nonumber\\
&&\ -\frac{1}{2\pi^2}\int_{0}^\infty\int_0^\infty\Big(\textrm{Re}\Big[\frac{e^{-i \phi_1 \ln K+i  \phi_2 \ln L}f(0;i\phi_1+1,-i\phi_2+1,-1,-1)}{ \phi_1 \phi_2}\Big]\nonumber\\
&&\ \ \ \ \  \ \ \ \ \ \ \ \ \ \ \ \ \ \ \ \ -\textrm{Re}\Big[\frac{e^{-i \phi_1 \ln K-i  \phi_2 \ln L}f(0;i\phi_1+1,i\phi_2+1,-1,-1)}{ \phi_1 \phi_2}\Big]\Big)\D \phi_1\D \phi_2,
\end{eqnarray}
and
\begin{eqnarray}\label{pi6}
 \Pi_{j,6}&:=&E\Big[e^{-\sum_{k=1}^{j} \Lambda(k)}V(j)I(-\ln V(j)> -\ln L, \ln S(T)\geq \ln K)\Big]\nonumber\\
 &=&E\Big[e^{-\sum_{k=1}^{j} \Lambda(k)+\ln V(j)}\Big]E^{\bar{Q}_3}\Big[I(-\ln V(j)>-\ln L, \ln S(T)\geq \ln K)\Big]\nonumber\\
  &=& \frac{1}{4}f(0;0,1,-1,-1)+\frac{1}{2\pi}\int_{0}^\infty \textrm{Re}\Big[\frac{e^{-i \phi_1 \ln K}f(0;i\phi_1,1,-1,-1)}{ i\phi_1 }\Big]\D \phi_1\nonumber\\
&&\ +\frac{1}{2\pi}\int_{0}^\infty \textrm{Re}\Big[\frac{e^{i \phi_2 \ln L}f(0;0,-i\phi_2+1,-1,-1)}{ i\phi_2 }\Big]\D \phi_2\nonumber\\
&&\ -\frac{1}{2\pi^2}\int_{0}^\infty\int_0^\infty\Big(\textrm{Re}\Big[\frac{e^{-i \phi_1 \ln K+i  \phi_2 \ln L}f(0;i\phi_1,-i\phi_2+1,-1,-1)}{ \phi_1 \phi_2}\Big]\nonumber\\
&&\ \ \ \ \ \ \ \ \ \ \ \ \ \ \ \ \ \ \ \ \ \ \ -\textrm{Re}\Big[\frac{e^{-i \phi_1 \ln K-i  \phi_2 \ln L}f(0;i\phi_1,i\phi_2+1,-1,-1)}{ \phi_1 \phi_2}\Big]\Big)\D \phi_1\D \phi_2,
\end{eqnarray}
where $\bar{Q}_3(O)$ is defined by
\begin{eqnarray*}
\bar{Q}_3(O)=\frac{E\Big[I(O)e^{-\sum_{k=1}^{j} \Lambda(k)+\ln V(j)}\Big]}{E\Big[e^{-\sum_{k=1}^{j} \Lambda(k)+\ln V(j)}\Big]},
\end{eqnarray*}
for any event $O\in \mathscr{F}_{T}$. Therefore, we have that
\begin{eqnarray*}\label{term 4}
&&E\Big[e^{-\sum_{k=1}^{j} \Lambda(k)}I( V(j)<L) V(j) (S(T)-K)^+ \Big] \nonumber\\
   &=&   E\Big[e^{-\sum_{k=1}^{j} \Lambda(k)+\ln S(T)} V(j)I(-\ln V(j)> -\ln L, \ln S(T)\geq \ln K)\Big] \nonumber\\
   && \ \ \ -K   E\Big[e^{-\sum_{k=1}^{j} \Lambda(k)} V(j)I( -\ln V(j)> -\ln L, \ln S(T)\geq \ln K)\Big]\nonumber\\
    &=& \Pi_{j,5} -K\Pi_{j,6}.
\end{eqnarray*}
Finally, we calculate $E\Big[e^{-\sum_{k=1}^{j-1} \Lambda(k)}I( V(j)<L) V(j)  (S(T)-K)^+ \Big]$ under $Q_4$ and $\bar{Q}_4$ defined by
\begin{eqnarray*}
Q_4(O)&=&\frac{E\Big[I(O)e^{-\sum_{k=1}^{j-1} \Lambda(k)+\ln S(T)+\ln V(j)}\Big]}{E\Big[e^{-\sum_{k=1}^{j} \Lambda(k)+\ln S(T)+\ln V(j)}\Big]},\\
\bar{Q}_4(O)&=&\frac{E\Big[I(O)e^{-\sum_{k=1}^{j-1} \Lambda(k)+\ln V(j)}\Big]}{E\Big[e^{-\sum_{k=1}^{j} \Lambda(k)+\ln V(j)}\Big]},
\end{eqnarray*}
for any event $O\in \mathscr{F}_{T}$. Following along similar arguments we obtain
\begin{eqnarray*}\label{term 5}
&&E\Big[e^{-\sum_{k=1}^{j-1} \Lambda(k)}I( V(j)<L) V(j) (S(T)-K)^+ \Big] \nonumber\\
   &=&   E\Big[e^{-\sum_{k=1}^{j-1} \Lambda(k)+\ln S(T)} V(j)I(-\ln V(j)> -\ln L, \ln S(T)\geq \ln K)\Big] \nonumber\\
   && \ \ \ -K   E\Big[e^{-\sum_{k=1}^{j-1} \Lambda(k)} V(j)I( -\ln V(j)> -\ln L, \ln S(T)\geq \ln K)\Big]\nonumber\\
    &=& \Pi_{j,7} -K\Pi_{j,8},
\end{eqnarray*}
where
\begin{eqnarray}\label{pi7}
 \Pi_{j,7}&:=&E\Big[e^{-\sum_{k=1}^{j-1} \Lambda(k)+\ln S(T)}V(j)I(-\ln V(j)> -\ln L, \ln S(T)\geq \ln K)\Big]\nonumber\\
  &=& \frac{1}{4}f(0;1,1,0,-1)+\frac{1}{2\pi}\int_{0}^\infty \textrm{Re}\Big[\frac{e^{-i \phi_1 \ln K}f(0;i\phi_1+1,1,0,-1)}{ i\phi_1 }\Big]\D \phi_1\nonumber\\
&&\ +\frac{1}{2\pi}\int_{0}^\infty \textrm{Re}\Big[\frac{e^{i \phi_2 \ln L}f(0;1,-i\phi_2+1,0,-1)}{ i\phi_2 }\Big]\D \phi_2\nonumber\\
&&\ -\frac{1}{2\pi^2}\int_{0}^\infty\int_0^\infty\Big(\textrm{Re}\Big[\frac{e^{-i \phi_1 \ln K+i  \phi_2 \ln L}f(0;i\phi_1+1,-i\phi_2+1,0,-1)}{ \phi_1 \phi_2}\Big]\nonumber\\
&&\ \ \ \ \ \ \ \ \ \ \ \ \ \ \ \ \ \ \ \ \  -\textrm{Re}\Big[\frac{e^{-i \phi_1 \ln K-i  \phi_2 \ln L}f(0;i\phi_1+1,i\phi_2+1,0,-1)}{ \phi_1 \phi_2}\Big]\Big)\D \phi_1\D \phi_2.
\end{eqnarray}
and
\begin{eqnarray}\label{pi8}
 \Pi_{j,8}&:=&E\Big[e^{-\sum_{k=1}^{j-1} \Lambda(k)}V(j)I(-\ln V(j)> -\ln L, \ln S(T)\geq \ln K)\Big]\nonumber\\
  &=& \frac{1}{4}f(0;0,1,0,-1)+\frac{1}{2\pi}\int_{0}^\infty \textrm{Re}\Big[\frac{e^{-i \phi_1 \ln K}f(0;i\phi_1,1,0,-1)}{ i\phi_1 }\Big]\D \phi_1\nonumber\\
&&\ +\frac{1}{2\pi}\int_{0}^\infty \textrm{Re}\Big[\frac{e^{i \phi_2 \ln L}f(0;0,-i\phi_2+1,0,-1)}{ i\phi_2 }\Big]\D \phi_2\nonumber\\
&&\ -\frac{1}{2\pi^2}\int_{0}^\infty\int_0^\infty\Big(\textrm{Re}\Big[\frac{e^{-i \phi_1 \ln K+i  \phi_2 \ln L}f(0;i\phi_1,-i\phi_2+1,0,-1)}{ \phi_1 \phi_2}\Big]\nonumber\\
&&\ \ \ \ \ \ \ \ \ \ \ \ \ \ \ \ \ \ \ \ \ \ \ -\textrm{Re}\Big[\frac{e^{-i \phi_1 \ln K-i  \phi_2 \ln L}f(0;i\phi_1,i\phi_2+1,0,-1)}{ \phi_1 \phi_2}\Big]\Big)\D \phi_1\D \phi_2.
\end{eqnarray}
Note that $\Pi_{j,7}$ and $\Pi_{j,8}$  can be obtained by replacing $f(0;\cdot,\cdot,-1,\cdot)$  in  $\Pi_{j,5}$ and $\Pi_{j,6}$  with $f(0;\cdot,\cdot,0,\cdot)$, respectively.
This completes the proof of the theorem.
\hfill$\Box$\\

\bibliographystyle{plain}
%
%\footnotesize

\end{CJK}
\end{document}